\documentclass[conference]{IEEEtran}
\IEEEoverridecommandlockouts
\usepackage{doublespace}

\usepackage[cmex10]{amsmath}
\usepackage{amssymb}
\usepackage{array}
\usepackage{url}

\usepackage{graphicx}
\usepackage{amsmath} 
\usepackage{cite}

	\newtheorem{theorem}{Theorem}

\newenvironment{example}{\par\begin{quotation}\small\noindent{\bf Example:\ }}{\end{quotation}\par}

\newenvironment{Proof}[1]{\medskip\par\noindent 
{\bf Proof:\, #1}\,}{{\mbox{\,$\bullet$}\par}}

	\newcommand{\thmref}[1]{(\ref{thm:#1})}
	\newcommand{\Thmref}[1]{Theorem~\ref{thm:#1}}
	\newcommand{\thmlabel}[1]{\label{thm:#1}}

	\newcommand{\twodef}[4]{\left\{\begin{array}{ll} 
		\displaystyle {#1} & {#2} \\
		\displaystyle {#3} & {#4}
		\end{array}\right.}
	\newcommand{\threedef}[6]{\left\{\begin{array}{ll} 
		\displaystyle {#1} & {#2} \\
		\displaystyle {#3} & {#4}\\
		\displaystyle {#5} & {#6}
		\end{array}\right.}

\newcommand{\imc}{{inscribed matter}}

\newcommand{\bstar}{{\beta^*}}

\newcommand{\bG}{{\bar{G}}}

\newcommand{\blankout}[1]{}

\newcommand{\be}{\begin{equation}}
\newcommand{\ee}{\end{equation}}
\newcommand{\bee}{\begin{IEEEeqnarray}{c}}
\newcommand{\eee}{\end{IEEEeqnarray}}
\newcommand{\bs}{\begin{IEEEeqnarray*}{c}}
\newcommand{\es}{\end{IEEEeqnarray*}}
\newcommand{\bd}{\begin{IEEEeqnarray*}{c}}
\newcommand{\ed}{\end{IEEEeqnarray*}}
\newcommand{\bea}{\begin{IEEEeqnarray}{rCl}}
\newcommand{\eea}{\end{IEEEeqnarray}}
\newcommand{\bas}{\begin{IEEEeqnarray*}{rCl}}
\newcommand{\eas}{\end{IEEEeqnarray*}}
\newcommand{\equat}[1]{equation (\ref{eq:#1})}
\newcommand{\equats}[1]{equations (\ref{eq:#1})}
\newcommand{\Equat}[1]{Equation (\ref{eq:#1})}

\newcommand{\xv}{{\bf x}}
\newcommand{\sv}{{\bf s}}

\newcommand{\tv}{{\bf t}}

\newcommand{\Hup}{{H^{\uparrow}}}
\newcommand{\tvv}{\vec{\tv}}
\newcommand{\tvec}{\vec{t}}
\newcommand{\Tvec}{{\vec{T}}}

\newcommand{\Svec}{\vec{S}}

\newcommand{\Svv}{\vec{\Smat}}
\newcommand{\Tvv}{\vec{\Tmat}}
\newcommand{\Xvv}{\vec{\Xmat}}

\newcommand{\bzero}{{\bf 0}}
\newcommand{\binfty}{{\boldsymbol{\infty}}}

\newcommand{\Dmat}{{\bf D}}

\newcommand{\Tmat}{{\bf T}}
\newcommand{\Smat}{{\bf S}}
\newcommand{\Xmat}{{\bf X}}

\newcommand{\mymu}{{\mu}}
\newcommand{\mylambda}{{\lambda}}

\newcommand{\myrho}{\rho}

\newcommand{\tbeta}{{\tilde{\beta}}}

\newcommand{\btau}{{\boldsymbol{\tau}}}

\author{Christopher Rose~\IEEEmembership{Fellow,~IEEE} and I.S. Mian}

\title{Inscribed Matter Communication: Part II}

\begin{document}

\maketitle

\begin{abstract}
This paper, combined with Part-I \cite{RoseMian16_1}, provides a comprehensive information-theoretic
treatment of the molecular communication problem at its finest grain -- emission and detection of
individual molecules, both identical (timing channel) or with embedded ``inscribed matter''
payloads.  Part-I provides the overarching framework while this, Part-II, develops results which
both extend previous timing channel results (``Bits Through Queues'' \cite{bits-Qs, sundaresan1,
  sundaresan2}) to emission schedules with deadlines as well as providing analytic expressions for a
quantity key to the analysis of the identical token timing channel -- the ``ordering entropy''
$H(\Omega|\Svv,\Tmat)$.  Expressions for $H(\Omega|\Svv,\Tmat)$ allow us to develop upper bounds for
the mutual information between input and output of the identical token timing channel which are then
used in Part-I to consider both the timing channel and the timing+payload channel.
\end{abstract}

\section{Introduction}
In Part-I {\cite{RoseMian16_1}} of this two-paper set we defined a signaling model and developed an
information theoretic framework for evaluating the capacity and efficiency of channels which use
molecules (or ``tokens'') as information carriers.  Here in Part-II we provide some necessary
undergirding results which are interesting in their own right.  In particular, we consider a timing
channel similar but not identical to Anantharam's and Verd\'u's ``Bits Through Queues'' channel
\cite{bits-Qs,sundaresan1, sundaresan2} wherein a mean launch time constraint is replaced by a
launch deadline constraint.  We derive closed forms for the optimizing distribution and the channel
capacity for this older timing channel and then apply the results to the molecular communication
problem.  We also derive analytic expressions and bounds for a key quantity in our analysis -- the
{\em ordering entropy} $H(\Omega|\Svv, \Tmat)$ -- first generally and then specifically for
exponential first-passage time distribution.  These results support the capacity bounds of Part-I
{\cite{RoseMian16_1}}, provide capacity results for a timing channel with an emission deadline
under exponential first-passage, and also establish that unlike the mean-constrained timing
channel, the worst case corruption is {\em not} exponential first-passage.  Our analysis ends with
the derivation of an upper bound on the token timing channel capacity.

\section{Brief Problem Description}
\label{sect:brief}
A detailed discussion of the underlying molecular communication problem and its importance in
both biology and engineering is provided in Part-I {\cite{RoseMian16_1}}.  Here we assume basic
familiarity with the concepts and provide only the mathematical description of the system.  As a reader aid,
key quantities are provided in TABLE~\ref{table:glossary} and in an identical table in Part-I.  
\begin{table}[h]
\begin{tabular}{p{2.0cm}|p{5.9cm}}  \hline
{\bf \em Token} &  {\small A unit released by the transmitter and captured by the receiver}\\ \hline
{\bf \em Payload} & {\small Physical information (\imc) carried by a token}\\ \hline
{\bf \em $\mylambda$} & {\small The average rate at which tokens are released/launched into the channel}\\ \hline
{\bf $\Tmat$} & {\small A vector of token release/launch times}\\  \hline
{\bf \em First-Passage} & {\small The time between token release/launch and token capture at the receiver}\\ \hline
{\bf $\Dmat$} & {\small A vector of first-passage times associated with launch times $\Tmat$}\\  \hline
{\bf $G(\cdot)$} & {\small The cumulative distribution function for first-passage random variable $D$}\\  \hline
{\bf $1/\mymu$} & {\small Average/mean first-passage time}\\  \hline
{\bf $\myrho$} & {\small $\mylambda/\mymu$, a measure of system token ``load'' }\\  \hline
{\bf $\Smat$} & {\small A vector of token arrival times, $\Smat = \Tmat + \Dmat$} \\  \hline
{\bf $P_k(\xv)$} & {\small A permutation operator which rearranges the order of elements in vector $\xv$}\\ \hline
{\bf $\Omega$} & {\small The ``sorting index'' which produces $\Svv$ from $\Smat$, {\em i.e.}, $\Svv = P_{\Omega}(\Smat)$}\\ \hline
{\bf $\Svv$} & {\small An {\em ordered} vector of arrival times obtained by sorting the elements of $\Smat$ (note, the receiver only sees $\Svv$ not $\Smat$)}\\  \hline
{\bf $I(\Smat;\Tmat)$} & {\small The mutual information between the launch times (input) and the arrival times (output)}\\  \hline
{\bf $I(\Svv;\Tmat)$} & {\small The mutual information between the launch times (input) and the {\em ordered} arrival times (output)}\\  \hline
{\bf $h(\Smat)$} & {\small The differential entropy of the arrival vector $\Smat$}\\  \hline
{\bf $H(\Omega|\Svv,\Tmat)$} & {\small The {\em ordering entropy} given the input $\Tmat$ and the output $\Svv$} \\  \hline
{\bf $\Hup(\Tmat)$} & {\small An upper bound for $H(\Omega|\Svv,\Tmat)$}\\  \hline
{\bf $C_q$ {\rm and } $C_t$} & {\small The asymptotic per token and per unit time capacity between input and output}\\  \hline
  \end{tabular}
$\mbox{ }$\\
\caption{Glossary of useful terms}
\label{table:glossary}
\end{table}

Thus, consider a communication system in which $M$ identical tokens are released/launched at times
$T_1, T_2, \cdots, T_M$ with no assumption that the $T_m$ are ordered in time. The duration of each
token's journey from transmitter to receiver is a random variable $D_m$ so that token $m$ arrives at
time $S_m = T_m + D_m$.  The $D_m$ are assumed independent and identically distributed (i.i.d.).  In
vector notation, we have $\Smat = \Tmat + \Dmat$.  We denote the density of each $D_m$ as
$f_{D_m}(d) = g(d)$, $d \ge 0$ and the cumulative distribution function (CDF) as $F_{D_m}(d) =
G(\cdot)$.  Likewise, the complementary CDF (CCDF) is $\bG(\cdot)$.  The channel output is the
time-sorted version of the $\{ S_m \}$ which we denote as $\{ \Svec_i \}$, $\Svec_i \le
\Svec_{i+1}$.

However, since the tokens are identical and their transit times are random, the receiver cannot
unequivocally know which arrival, $\Svec_i$ corresponds to which transmission $T_m$.  That is,
$\Svv$, the ordered arrival times are related to $\Smat$ through a permutation operation,
$P_{\Omega}(\Svv) = \Smat$ and from the receiver's perspective, $\Omega$ is a random variable,
$\Omega = 1, 2, \cdots, M!$.

In the next section, we provide a sampling of results from Part-I upon which we will expand here in
Part-II.

\section{Key Results from the Companion Paper \cite{RoseMian16_1}}
A good deal of effort was expended in Part-I quantifying the relationships between $\Tmat$, $\Dmat$,
$\Smat$ and $\Svv$, and in developing a signaling discipline wherein the measure of communication
efficacy is determined by the mutual information between $\Svv$ and $\Tmat$, $I(\Svv;\Tmat)$. That
is, we took care to make sure that channel coding theorem results \cite[(chapt 8 \& 10)]{cover}
could be applied by deriving a model in which channel uses were (asymptotically) independent.

Specifically, we assume sequential finite signaling intervals/epochs of duration $\tau$ and then
define the token intensity as $\mylambda = \frac{M}{\tau}$ as a proxy for transmitter power (each
emission ``costs'' some fixed energy).  In addition, we assume that the {\em mean first-passage
  time} exists with $E[D] = 1/\mymu$ so that tokens always (eventually) arrive at the receiver.  It
is important to note that finite first-passage time is important for information-theoretic patency
of the analysis.  As shown in Part-I \cite{RoseMian16_1}, finite first-passage allows sequential
signaling intervals (channel uses) to be derived which are, in the limit of long intervals,
asymptotically independent.  Infinite first-passage {\em does not allow such asymptotically
  independent sequential intervals to be constructed} so that mutual information $I(\Svv;\Tmat)$ is
not necessarily the proper measure of information carriage for the system.

We note that transport processes such as free-space diffusion do {\em not} have finite
first-passage.  However, any physical system is limited in extent and therefore does have finite
(though perhaps long) first-passage under an ergodic transport model.  So, the analysis holds for
situations where tokens eventually arrive at the receiver. Of course, as discussed in Part-I, there
are situations where a token might never arrive at {\em any} time.  Such situations include channels
where the token ``denatures'' and becomes unrecognizable by the receiver or is ``gettered'' by
agents in the channel which remove the token from circulation before detection \cite{farsad_isit16,
  farsadIT16}.  Such tokens do not contribute to intersymbol interference (earlier tokens
corrupting a subsequent interval) so it is possible that slightly different first-passage time
distributions could be used which still preserve asymptotically independent channel uses.  However,
since any such model produces a first passage density, $g(d)$ with singularities, the specific
analysis used in Part-I is not immediately applicable. The implications (and shortcomings) of the
finite first-passage assumption are discussed more carefully in the Discussion \& Conclusion section
of Part-I.

Now, as a prelude to deriving channel capacity, we recall from Part-I \cite{RoseMian16_1} that if
$Q(\xv)$ is a hypersymmetric function, $Q(\xv) = Q(P_k(\xv))$ $\forall k$ where $P_k(\cdot)$ is a
permutation operator and $\Xmat$ is a hypersymmetric random vector whose PDF obeys $f_{\Xmat}(\xv) =
f_{\Xmat}(P_k(\xv))$, then when $\vec{\Xmat}$ is the ordered version of random vector $\Xmat$ we
have
\bee
\label{eq:hyperexpect}
E_{\Xvv} \left [Q(\Xvv) \right ]
=
E_{\Xmat} \left [Q(\Xmat) \right ]
\eee
This expression (Theorem 1 from Part-I) allows us to avoid deriving order distributions on
potentially correlated random variables.

Next, the mutual information between the input $\Tmat$ and the output $\Svv$ of the token timing
channel is given by
\bee
I(\Svv; \Tmat)
=
h(\Svv) - h(\Svv|\Tmat)
\label{eq:ISvvT}
\eee
Then, if we assume that $g(\cdot)$ does not contain singularities, we observe that the set of all $\Smat$ for which two or more elements are equal is of zero measure which allows us to ``fold'' the distribution on $f_{\Smat}(\cdot)$ to obtain a distribution on the ordered $\Svv$.  If we then in addition
assume hypersymmetric $\Xmat$, we can write \equat{ISvvT} as
\bee
I(\Svv; \Tmat)
=
h(\Smat) - \log M! - h(\Svv|\Tmat)
\label{eq:IST_SvvT}
\eee
Hypersymmetry of $\Xmat$ and no singularity in $g(\cdot)$ implies that we can ignore
situations where one or more of the $S_i$ are equal, which then implies an equivalence
\bee
\{ \Svv, \Omega \} \Leftrightarrow \Smat
\label{eq:equivalence}
\eee
which leads to
\bee
h(\Smat|\Tmat) = h(\Svv, \Omega|\Tmat) = H(\Omega|\Svv,\Tmat) + h(\Svv|\Tmat)
\label{eq:hSvvTequiv}
\eee
where $H(\Omega|\Svv,\Tmat)$ is the {\em ordering entropy}, a measure of the uncertainty about which
$S_m$ correspond to which $\Svec_i$.  \Equat{hSvvTequiv} allows us to write the \equat{IST_SvvT} as
\be
\label{eq:orderedMI_decomp}
I(\Svv; \Tmat)
=
I(\Smat;\Tmat)
-
\left ( \log M! - H(\Omega|\Svv,\Tmat) \right )
\ee
And since we know asymptotically independent channel uses can be assured (Part-I Theorem 2,
\cite{RoseMian16_1}), the channel capacity in bits/nats per channel use is
\bd
C
=
\max_{f_{\Tmat}(\cdot)}
\left [
I(\Smat;\Tmat)
-
\left ( \log M! - H(\Omega|\Svv,\Tmat) \right )
\right ]
\ed

We then derived an upper bound for the ordering entropy $H(\Omega|\Svv,\Tmat)$ in Part-I as
\bee
\label{eq:HOHupineq}
H(\Omega|\Svv,\tv) \le \Hup(\tv)
\eee
and derived/defined $\Hup(\cdot)$ as
\bea
\IEEEeqnarraymulticol{3}{l}{\Hup(\tv) = \sum_{\ell=1}^{M-1} \log(1 + \ell)} \nonumber \\ \quad
& \times & 
{\sum_{m=\ell}^{M-1}}
{ \sum_{|\bar{\xv}| = \ell}} { \prod_{j=1}^m 
\bG^{\bar{x}_j}(\tvec_{m+1} - \tvec_j)
G^{1 -\bar{x}_j}(\tvec_{m+1} - \tvec_j)}  \IEEEeqnarraynumspace
\label{eq:Hexpurg3}
\eea
with $\tvv$ the size-ordered version of $\tv$ and $\bar{\xv}$ a binary $m$-vector with $|\bar{\xv}|$
defined as its number of non-zero entries.  Then, through hypersymmetry arguments as in
\equat{hyperexpect}, we showed that
\bee
H(\Omega|\Svv,\Tmat) = E_{\tv} \left [ H(\Omega|\Svv,\tv) \right ]  \le E_{\tvv} \left [ \Hup(\tv) \right ] = \Hup(
\Tmat) \IEEEeqnarraynumspace
\label{eq:HOinequality}
\eee
with equality {\bf iff} the first passage time is exponential (Theorem 8, Part-I \cite{RoseMian16_1}).

Based on asymptotically independent channel uses, two key measures of channel capacity were
derived. The first, $C_q$, is the asymptotic per token capacity:
\bee
C_q
=
\lim_{M \rightarrow \infty}
\frac{1}{M} I(\Svv;\Tmat)
\label{eq:Cqdef}
\eee
and the second is $C_t = \mylambda C_q$, the asymptotic per unit time capacity
(Theorem 4, Part-I \cite{RoseMian16_1}).

In what follows we will seek to maximize $h(\Smat)$ under the deadline constraints on the $\Tmat$,
derive a variety of expressions for $\Hup(\Tmat)$ for general and then for exponential first-passage
under both a deadline and also the mean launch constraint considered in ``Bits Through Queues''
\cite{bits-Qs} and elsewhere \cite{sundaresan1, sundaresan2}.  We follow with asymptotic results for
$\Hup(\Tmat)/M$ as $M \rightarrow \infty$ again assuming exponential first-passage and close by
providing upper bounds for $C_q$ and $C_t$.

\section{``Bits Through Queues'' With a Deadline Constraint}
\label{sect:minmaxhS}
\subsection{Preliminaries}
The award-winning paper, ``Bits Through Queues'' \cite{bits-Qs} and others
\cite{sundaresan1,sundaresan2} derived capacity results for a timing channel under a mean launch
time constraint.  In this section we derive results for a similar single-token timing channel where
instead of a mean constraint, the launch time $T$ is limited to $[0,\tau]$ \cite{isit11,SonRos13}
and first-passage is exponential with parameter $\mymu$.  Here we provide closed forms for both the
capacity and for the capacity-achieving input density. However, unlike in \cite{bits-Qs} we show
that exponential first-passage is {\em not} the worst case corruption for the launch
deadline-constrained channel.

Since $T$ is independent of $D$, the density of $S=T+D$ is given by
\bd
f_{S}(s) = \int_0^s f_T(t) f_D(s-t) dt \quad 0 \leq s 
\ed
and because $T$ is constrained to $[0,\tau]$, we can divide $f_S(s)$ into two regions:
region $I$ where $s \in [0,\tau]$ and region $II$ where $s \in (\tau,\infty)$.  We then have
\bee
f_S(s)= \left\{
\begin{array}{l l}
\sigma f_{S|I}(s)  & 0 \leq s \leq \tau\\
(1-\sigma) f_{S|II}(s)  & s > \tau\\
\end{array} \right.
\label{eq:pdf_S1S2}
\eee
where
\bd
\sigma
=
\int_{0}^{\tau} f_{S}(s) ds
\ed
with
\bee
\sigma f_{S|I}(s)
=
\int_0^s f_T(t) f_D(s-t) dt 
\eee
and
\bee
(1- \sigma) f_{S|II}(s) 
=
\int_{0}^{\tau} f_T(t) f_D(s-t) dt
\eee

For $D$ exponential with parameter $\mymu$ we have
\bee
\label{eq:pdf_S}
f_{S}(s) = \int_0^s f_T(t) \mymu e^{-\mymu(s-t)} dt \quad 0 \leq s 
\eee
and
\bee
\sigma f_{S|I}(s)
=
\int_0^s f_T(t) \mymu e^{-\mymu(s-t)} dt
\eee
and
\bee
\label{eq:fSII}
(1- \sigma) f_{S|II}(s) 
=
e^{-\mymu s}
\int_{0}^{\tau} f_T(t) \mymu e^{\mymu t} dt 
\eee
The entropy of $S$ is then
\bee
\label{eq:entropy_S}
\begin{split}
h(S) & = -\int_0^{\infty} f_{S}(s) \log f_{S}(s) ds \\
     & = -\int_0^{\tau} \sigma f_{S|I}(s) \log \left ( \sigma f_{S|I}(s) \right ) ds\\
     & - \int_{\tau}^{\infty} (1-\sigma )f_{S|II}(s) \log \left ( (1-\sigma )f_{S|II}(s)\right ) ds\\
     & =  \sigma h(S|I) + (1-\sigma) h(S|II) + H_B (\sigma) \\
\end{split}
\eee
where $H_B(\cdot)$ is the binary entropy function.  Notice that no particular care
has to be taken with the integrals at $s=T$ because $f_S(s)$ cannot contain
singularities -- it is obtained by the convolution of two densities, one of
which, $f_D(\cdot) = g(\cdot)$, contains no singularities.

\subsection{Maximization of $h(S)$}
\label{sect:optimize}

\begin{figure}[h]
\centering
\includegraphics[width=3in]{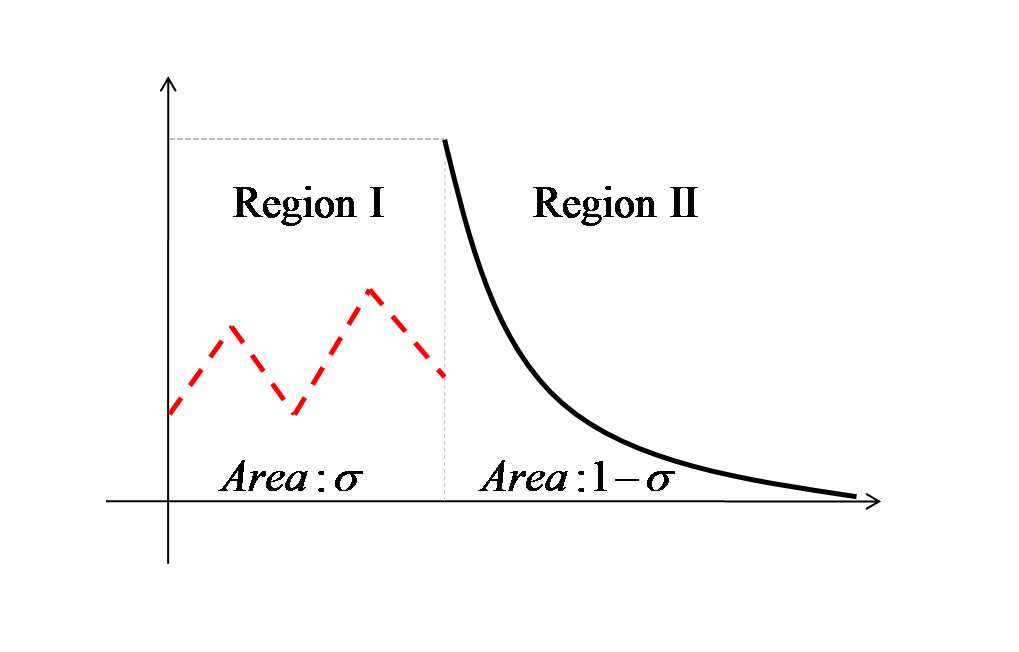}
\caption{The shapes associated with $f_{S}(s)$: We assume arbitrary shape in region I and the requisite exponential shape in region II.}
\label{fig:step0}
\end{figure}

We observe of \equat{fSII} that the {\em shape} of the
conditional density for $s>\tau$ is completely determined -- an exponential with
parameter $\mymu$ as depicted in FIGURE~\ref{fig:step0}.
Thus, selection of $f_T(\cdot)$ does not affect $f_{S|II}(\cdot)$ and
we must have $h(S|II) = 1 - \log \mymu$.

This observation suggests a three-step approach to maximizing $h(S)$.  In the
first two steps, we completely ignore $f_T(\cdot)$ and find the {\em shape} $f_{S|I}(\cdot)$
and value of $\sigma$ which maximize \equat{entropy_S}.  In step three, we determine that
there indeed exists a density $f_T(\cdot)$ which produces the optimizing $f_S(\cdot)$.

\begin{figure}[h]
\centering
\includegraphics[width=3in]{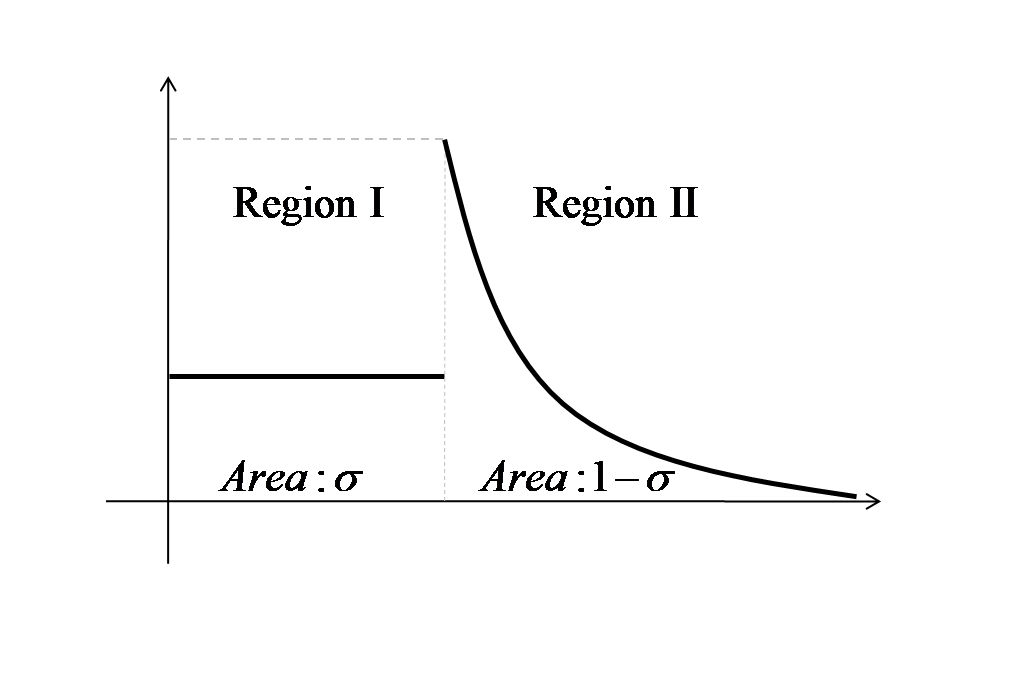}
\caption{The updated shape of $f_{S}(s)$ after step 1: $f_{S|I}(s)$ is chosen as $\frac{1}{\tau}$. }
\label{fig:step1}
\end{figure}

{\bf Step 1:} For fixed $\sigma$ we see from \equat{entropy_S} that $h(S)$ is
maximized solely by our choice of $f_{S|I}(\cdot)$. The uniform density maximizes
entropy on a finite interval \cite{Cover06}.  Thus, $f_{S|I}(s) = \frac{1}{\tau}$
and $h(S|I) = \log \tau$ as depicted in FIGURE~\ref{fig:step1}.

{\bf Step 2:} Since for any $\sigma$, $h(S|I) = \log \tau$, we have
\be
h(S)
=
\sigma \log \tau + 
(1-\sigma) (1 - \log \mymu) + H_B (\sigma) \\
\label{eq:HSstep1}
\ee
Taking the derivative of \equat{HSstep1} with respect to $\sigma$ yields
\bd
\log \tau
-
(1-\log \mymu)
-
(1+\log \sigma)
+
(1+\log (1-\sigma))
\ed
which we set to zero to obtain
\bd
\log \mymu \tau
-
\log \frac{\sigma}{1-\sigma}
-
1
=
0
\ed
We rearrange to obtain
\bd
\mymu \tau
=
\frac{e\sigma}{1-\sigma}
\ed
from which we deduce that the optimal $\sigma$ is
\bee
\label{eq:sigmastar}
\sigma^*
=
\frac{\mymu \tau}{e  + \mymu \tau}
\eee

Returning to the entropy maximization we have
\bd
\max_{f_T(\cdot)} h(S)
\le
\sigma^* \log \tau
+
(1-\sigma*) (1 - \log \mymu) + H_B (\sigma^*)
\ed
which through substitution of $\sigma^*$ according to \equat{sigmastar}
yields
\bee
\max_{f_T(\cdot)} h(S)
\le
\log \left ( \frac{e + \mymu \tau}{\mymu} \right )
\label{eq:hSmax}
\eee
with equality when
\bee
f_S(s)
=
\twodef{\frac{\mymu}{e + \mymu \tau} }{0 \le s < \tau}{\frac{e}{e+\mymu \tau} \mymu e^{- \mymu (s-\tau)}}{s \ge \tau}
\label{eq:optimumS}
\eee
\blankout{
as depicted in FIGURE~\ref{fig:step2}.
\begin{figure}[h]
\centering
\includegraphics[width=3in]{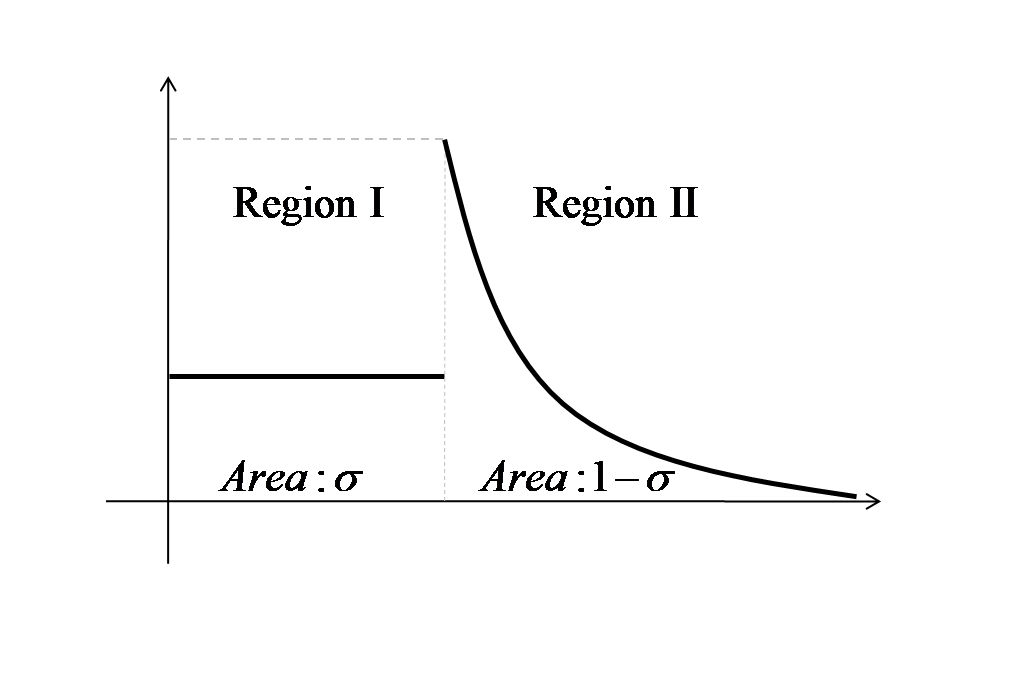}
\caption{The optimizing $f_S(s)$.}
\label{fig:step2}
\end{figure}}

{\bf Step 3:} All that remains is to ascertain whether $\exists f_T(\cdot)$ which
can generate the $f_S(s)$ of \equat{optimumS}.  Since $f_S(\cdot)$ is the convolution of
$f_D(\cdot)$ and $f_T(\cdot)$ we can use Fourier transforms to obtain a candidate solution for $f_T(\cdot)$.  That is, the Fourier transform of $f_D(\cdot)$ is $\frac{\mymu}{\mymu + j 2 \pi f}$ so the Fourier transform of $f_T(\cdot)$ is
\bd
{\cal F} \left  \{ f_T(\cdot) \right \}
=
{\cal F} \left  \{ f_S(\cdot) \right \}
\left (
\frac{j 2 \pi f}{\mymu} + 1
\right )
\ed
Multiplication by $j 2 \pi f$ implies differentiation so we must have
\bd
f_T(t)
=
\frac{1}{\mymu}
\frac{d}{dt} f_S(t)
+
f_S(t)
\ed
which implies via \equat{optimumS} that
\bee
\label{eq:optimalfT}
f_T(t)
=
\twodef{\frac{\mymu}{e+\mymu \tau}}{0<t<\tau}{\delta(t) \frac{1}{e + \mymu \tau} + \delta(t-\tau) \frac{e-1}{e + \mymu \tau}}{\mbox{o.w.}}
\eee
-- a valid probability density function.

We can now state the maximum mutual information (capacity in bits per channel use) as
\bee
\label{eq:maxI}
\max_{f_T(\cdot)}
I(S;T)
=
\log \left ( \frac{e  \mymu \tau}{\mymu} \right ) - (1- \log \mymu)
=
\log \left ( 1 + \frac{\mymu \tau}{e} \right ) \IEEEeqnarraynumspace
\eee
which is achieved using the emission time density of \equat{optimalfT}.

We summarize the result as a theorem:
{\em \begin{theorem} {\bf Maximum \boldmath $I(S;T)$ Under a Deadline Constraint:}
\thmlabel{fTopthSopt}

If $S=T+D$ where $D$ is an exponential random variable with parameter $\mymu$ and $T \in [0,\tau]$, then the
mutual information between $S$ and $T$ obeys
\bd
I(S;T) \le \log \left ( 1 + \frac{\mymu \tau}{e} \right )
\ed
with equality when 
\bd
f_T(t)
=
\twodef{\frac{\mymu}{e+\mymu \tau}}{0<t<\tau}{\delta(t) \frac{1}{e + \mymu \tau} + \delta(t-\tau) \frac{e-1}{e + \mymu \tau}}{\mbox{o.w.}}
\ed
and
\bd
f_S(s)
=
\twodef{\frac{\mymu}{e + \mymu \tau} }{0 \le s < \tau}{\frac{e}{e+\mymu \tau} \mymu e^{- \mymu (s-\tau)}}{s \ge \tau}
\ed
\end{theorem}}
\begin{Proof}{Theorem \thmref{fTopthSopt}}
See the development leading to the statement of \equat{maxI}.
\end{Proof}

The only remaining question is whether for interval-limited inputs, the exponential first-passage
time density, to quote \cite{bits-Qs} ``plays the same role ... that Gaussian noise plays in
additive noise channels.''  Unfortunately the answer is no, a result we state as a theorem:
{\em \begin{theorem} {\bf \boldmath For $T$ Constrained to $[0,\tau]$, the Minmax Mutual Information First-Passage Density Is NOT Exponential:}
\thmlabel{minmax}

If $g(\cdot)$ is a first passage density with mean $1/\mu$ and
$f_T(\cdot)$ can be nonzero only on $[0,\tau]$, then
\bd
\operatorname*{arg\,min}_{g(\cdot)}
\left [ \max_{f_T(\cdot)}
I(S;T)
\right ]
=
g^*(s)
\ne
\mu e^{-\mu s} u(s)
\ed
where $u(\cdot)$ is the unit step function.
\end{theorem}}
\begin{Proof}{Theorem \thmref{minmax}}
Consider that
\bee
I(S;T) = \int \int  f_T(t) g(s-t) \frac{g(s-t)}{f_S(s)} dt ds
\label{eq:mutualinfodef}
\eee
is convex in $g(\cdot)$ \cite{cover,gallagerit}.  Since we constrain $g(\cdot)$ to be non-negative with mean
$1/\mu$ and unit integral, we can apply Euler-Lagrange variational techniques \cite{hild}.  That is, we
set $q(x) = g(x) + \epsilon \eta(x)$ where $\eta(x)$ is any function defined on $[0,\infty)$, and
look for the stationary point
\bea
\IEEEeqnarraymulticol{3}{l}{\frac{d}{d\epsilon} \left [ \int \int  f_T(t) q(s-t) \frac{ q(s-t) }{\int f_T(x) q(s-x) dx}
dt ds \right .}  \nonumber \\
& + &  \left .  a \left ( \int s q(s) ds - \frac{1}{\mu}  \right )
+ b \left (\int q(s) ds - 1 \right ) 
\right ]_{\epsilon = 0}
=0 \IEEEeqnarraynumspace
\label{eq:eulerlagrange}
\eea
where $a$ and $b$ are (Lagrange) multipliers.
Satisfaction of \equat{eulerlagrange} for any possible $\eta(\cdot)$ requires (after expansion and a change of coordinate systems in the double integral) that
\bee
\log g(s)
=
\int_0^\tau f_T(t) \log f_S(s+t) dt + a s + b
\label{eq:variationalg}
\eee
for the $g(\cdot)$ that minimizes \equat{mutualinfodef}.

Now, from \Thmref{fTopthSopt} we know the form of the optimizing $f_T(t)$, $t \in [0,\tau]$ and the
resulting $f_S(s)$ were $g(\cdot)$ exponential with mean $1/\mu$.  We also know that $I(S;T)$ is
concave in $f_T(t)$\cite{cover,gallagerit}.  Thus, were exponential $g(\cdot)$ to minimize the
maximum mutual information, the left hand side of \equat{variationalg} would be a linear function of
$s$.  Thus, the integral term on the right would also need to be a linear function of $s$ given
$f_S(s)$ as in \equat{optimumS} and $f_T(t)$ as in \equat{optimalfT}.

For $s \ge \tau$ we have $f_S(s+t) = \frac{\mu e}{e + \mu \tau} e^{ - \mu (s+t-\tau)}$ and thence
\bea
\IEEEeqnarraymulticol{3}{l}{\int_0^\tau f_T(t) \log f_S(s+t) dt} \nonumber \\ \quad
 & =& 
\int_{0}^{\tau}
\frac{
\left (
\delta(t)
+
\mu
+
\delta(t-\tau)(e-1)
\right )
(-\mu (s+t-\tau))}{e + \mu \tau} dt \nonumber \\
 & + & \log\frac{\mu e}{e + \mu \tau}   \nonumber \\
 & = & 
\frac{\mu \tau - \mu s e}{e + \mu \tau}
-
\frac{\mu^2}{e + \mu \tau}
\left . \frac{(s+t - \tau)^2}{2} \right |_0^\tau 
 + \log\frac{\mu e}{e + \mu \tau}   \nonumber \\
 & = & \log\frac{\mu e}{e + \mu \tau}
 +\frac{\mu \tau - \mu s e}{e + \mu \tau}
-
\frac{\mu^2}{e + \mu \tau}
\left ( s \tau - \frac{\tau^2}{2} \right )
\eea
which is indeed a linear function of $s$.

However, when $0 \le s < \tau$ we obtain
\bea
\IEEEeqnarraymulticol{3}{l}{\int_0^\tau f_T(t) \log f_S(s+t) dt} \nonumber \\ \quad
 & =& 
\int_0^{\tau -  s}
\left (\frac{\delta(t)}{e +  \mu\tau} 
 +  \frac{\mu}{e +  \mu \tau} \right )
 \log  \frac{\mu}{e + \mu \tau}
dt \nonumber \\
 & + & 
\int_{\tau - s}^{\tau}
\left ( \frac{\mu}{e + \mu \tau}
+  \frac{\delta(t-\tau)(e-1)}{e + \mu \tau} 
\right )
 \log  \frac{\mu e}{e + \mu \tau} dt \nonumber \\
 & - & 
\int_{\tau - s}^{\tau}
\left ( \frac{\mu}{e + \mu \tau}
+  \frac{\delta(t-\tau)(e-1)}{e + \mu \tau} 
\right )
\mu (s+t-\tau) dt \nonumber \\
 & =& 
\frac{1 + \mu (\tau - s)}{e + \mu \tau}
 \log  \frac{\mu}{e + \mu \tau}
+
\frac{\mu s +e - 1}{e + \mu \tau}
 \log \frac{\mu e}{e + \mu \tau} \nonumber \\
 & - & 
\frac{\mu s (e-1)}{e + \mu \tau}
-
\frac{\mu^2}{e + \mu \tau}
\left . \frac{(s+t-\tau)^2}{2}\right  |_{\tau-s}^\tau \nonumber \\
 & = & 
 \log  \frac{\mu}{e + \mu \tau}
+
\frac{2\mu s + e(1- \mu s) - 1}{e + \mu \tau}
-
\frac{\mu^2}{e + \mu \tau}
\frac{s^2}{2} \IEEEeqnarraynumspace
\eea
which does not have the requisite form owing to the term in $s^2$.

Therefore, for $T$ constrained to $[0,\tau]$, the minmax $I(S;T)$ first-passage density, $g(\cdot)$, is not
exponential.
\end{Proof}
It is important to note that owing to a faulty proof \cite{isit11}, exponential first passage was
previously claimed to maximally suppress capacity of the constrained-launch channel.
\Thmref{minmax} corrects this error.

\section{Ordering Entropy, $H(\Omega|\Svv,\Tmat)$}
In this section we derive a number of results for the ordering entropy, $H(\Omega|\Svv,\Tmat)$, both
generally and for exponential first-passage.  As a prelude, we recall from section~\ref{sect:brief}
and TABLE~\ref{table:glossary} that $\tvv = \{\tvec_1, \tvec_2, \cdots, \tvec_M \}$ is the ordered
version of $\tv = \{t_1, t_2, \cdots, t_M\}$, the launch times, and that $G(\cdot)$ is the
cumulative distribution function (CDF) for the first-passage time $D$ (with $\bG(\cdot)$ its
complementary cumulative distribution function (CCDF).  We recall from Part-I \cite{RoseMian16_1}
that $H(\Omega|\Svv,\tv) \le \Hup(\tv)$ from \equat{HOHupineq} with $\Hup(\tv)$ defined as in
\equat{Hexpurg3}. We also recall from Part-I that $\bar{\xv}$ is a binary vector of dimension $m$
and $\sum_{|\bar{\xv}| = \ell}$ is a sum over all $\bar{\xv}$ containing exactly $\ell$ $1$'s.  The
inequality in \equat{HOinequality} is an equality {\bf iff} first-passage is exponential with $\bG(x) =
e^{-\mymu x} u(x)$, where $u(x)$ is the unit step function (Theorem 9, Part-I).

\subsection{General Calculation of $\Hup(\tv)$}
\label{app:Hup}
To calculate $\Hup(\tv)$ we first define
\bee
\label{eq:thetadef}
\Theta_{m,\ell}(\tvv)
\equiv
\sum_{|\bar{\xv}| = \ell}
\prod_{j=1}^m
\bar{G}^{\bar{x}_j}(\tvec_{m+1} - \tvec_j)
G^{1 -\bar{x}_j}(\tvec_{m+1} - \tvec_j) \IEEEeqnarraynumspace
\eee
which implies via \equat{Hexpurg3} that
\bee
\label{eq:Hexpurg4}
\Hup(\tv)
=
\sum_{\ell=1}^{M-1}
\log(1 + \ell)
\sum_{m=\ell}^{M-1}
\Theta_{m,\ell}(\tvv)
\eee
\blankout{Please note that we have assumed that $t_1 \le t_2 \le \cdots \le t_m$ -- important
because although $\left | \Omega \right |_{\sv,\tv}$ is
hypersymmetric in $\tv$, $\Hup(\tv)$ is {\em not}.}

In principle, we could derive $\Hup(\Tmat)$ by taking the expectation of \equat{Hexpurg4} with
respect to ordered emission times, $\tvv$.  However, direct {\em analytic} evaluation of $\Hup(\Tmat)$
requires we derive joint order densities on the underlying $\Tmat$, a difficult task in general when
the individual $\{T_m \}$ are not necessarily independent.

So, we take a different approach.  The sum over all permutations of binary vector $\bar{\xv}$ in the
definition of $\Theta_{m,\ell}(\tvv)$ (\equat{thetadef}) renders it hypersymmetric in $\tvec_1 ,
\cdots , \tvec_m$ given the $(m+1)^{\mbox{st}}$ smallest emission time $\tvec_{m+1}$.  That is,
$\Theta_{m,\ell}(\tvv) = \Theta_{m,\ell}(P_k(\tvec_1, \cdots,\tvec_m), \tvec_{m+1})$ for any
permutation function $k$ so long as $\tvec_{m+1}$ is fixed. In what follows we therefore drop the
over-vector notation for the $t_1, t_2, \cdots, t_m$ and assume all are less than $\tvec_{m+1}$.

Therefore, by \equat{hyperexpect} we can define  $E[\Theta_{m,\ell}] = \bar{\Theta}_{m,\ell}$ as
\bee
E_{\Tvec_{m+1}} \left   [
E_{T_1,\cdots,T_m|\Tvec_{m+1}}
\left [
\Theta_{m,\ell}(T_1,\cdots,T_m,\Tvec_{m+1})
\right ]
\right ] \IEEEeqnarraynumspace
\label{eq:Hhyper}
\eee
Then, the CDF, $F_{\vec{T}_{m+1}} (t_{m+1})$, of the $(m+1)^{\mbox{st}}$ smallest emission time is
\bea
\IEEEeqnarraymulticol{3}{l}{F_{\vec{T}_{m+1}} (t_{m+1})} \nonumber \\ \quad
 & = &
1
-
\displaystyle \sum_{k=0}^{m}
{M \choose k}
\underbrace{
\int_{\bzero}^{\bf t_{m+1}}
}_{\mbox{$k$}}
\underbrace{
\int_{\bf t_{m+1}}^{\binfty}
}_{\mbox{$M-k$}}
f_{\Tmat}(\tv)
dt_M 
\cdots
dt_{1} \IEEEeqnarraynumspace
\label{eq:ftvm}
\eea
and likewise, the CDF, $F_{T_1,\cdots,T_m|\Tvec_{m+1}} (t_1,\cdots,t_m|\tvec_{m+1})$,  of the smallest {\em unordered} $T_1,\cdots,T_m$ given $\vec{T}_{m+1}$ is
\bea
\IEEEeqnarraymulticol{3}{l}{F_{T_1,\cdots,T_m|\Tvec_{m+1}} (t_1,\cdots,t_m|\tvec_{m+1})}  \nonumber \\ \quad
 &  = & \frac{F_{T_1,\cdots,T_m} (t_1,\cdots,t_m)}{F_{T_1,\cdots,T_m} (\tvec_{m+1},\cdots,\tvec_{m+1})}
\eea
$\forall t_j \le \vec{t}_{m+1}$ where $j=1,\cdots,m$.

Therefore, by the hypersymmetry of ${\Theta}_{m,\ell}$ in $t_1,\cdots,t_m$ we may write $\bar{\Theta}_{m,\ell}$ as
\bee
\int_0^{\infty}
{\bf \int}_{\bzero}^{{\bf t_{m+1}}}
\frac{f_{\vec{T}_{m+1}}(t_{m+1}) f_{{\bf T_m}}({\bf t_m})B(m,\ell,\tv)}{F_{{\bf T_m}}(t_{m+1},\cdots,t_{m+1})}
d{\bf t_m}
dt_{m+1}   \IEEEeqnarraynumspace
\label{eq:Thbardef}
\eee
where  ${\bf T_m} = \{T_1, \cdots, T_M \}$, ${\bf t_m} = \{T_1, \cdots, t_M \}$  and
\bee
\label{eq:Bdef}
B(m,\ell,\tv)
\equiv
{m \choose \ell}
\!\prod_{j=1}^{\ell}
\!\bar{G}(t_{m+1} \!-\! t_j)\!\!\prod_{k=\ell+1}^m
\!G(t_{m+1} \!-\! t_k) \IEEEeqnarraynumspace
\eee
and thence
\be
\label{eq:He1}
\Hup(\Tmat)
=
\sum_{\ell=1}^{M-1}
\log(1 + \ell)
\sum_{m=\ell}^{M-1}
\bar{\Theta}_{m,\ell}
\ee
In addition, if we define
\be
\label{eq:Gamma}
\Gamma_{M,\ell}
=
\sum_{m=\ell}^{M-1}
\bar{\Theta}_{m,\ell}
\ee
and
\be
\label{eq:DeltaGamma}
\Delta\Gamma_{M,\ell}
=
\Gamma_{M,\ell} -  \Gamma_{M,\ell+1}
\ee
then we can also express $\Hup(\Tmat)$ as
\be
\label{eq:He2}
\Hup(\Tmat)
=
\sum_{\ell=1}^{M-1}
\Delta\Gamma_{M,\ell}
\log (\ell + 1)!
\ee
The development starting in section~\ref{app:Hup} proves the following theorem:
{\em \begin{theorem}{\bf The General Form of $\Hup(\Tmat)$:}
\thmlabel{HupGeneral}

If we define 
$$
\Gamma_{M,\ell}
=
\sum_{m=\ell}^{M-1}
\bar{\Theta}_{m,\ell}
$$
and
$$
\Delta\Gamma_{M,\ell}
=
\Gamma_{M,\ell} -  \Gamma_{M,\ell+1}
$$
where  $\bar{\Theta}_{m,\ell}$ is as defined by \equats{Thbardef} and \equat{Bdef}.
\blankout{\bd
\begin{array}{l}
\bar{\Theta}_{m,\ell}
=
\displaystyle \int_0^{\infty}
f_{\vec{T}_{m+1}}(t_{m+1}) \times\\

\displaystyle {\bf \int}_{\bzero}^{{\bf t_{m+1}}}
\left [
  \begin{array}{c}
\displaystyle \frac{f_{T_1,\cdots,T_m}(t_1,\cdots,t_m)}{F_{T_1,\cdots,T_m}(t_{m+1},\cdots,t_{m+1})}\\
{\mbox{\Large \boldmath $\times$} }\\
B(m,\ell,\tv)
\end{array}
\right ]
dt_1
\cdots
dt_{m+1}
\end{array}
\ed}
\blankout{
\bee
\int_0^{\infty}
{\bf \int}_{\bzero}^{{\bf t_{m+1}}}
\frac{f_{\vec{T}_{m+1}}(t_{m+1}) f_{{\bf T_m}}({\bf t_m})B(m,\ell,\tv)}{F_{{\bf T_m}}(t_{m+1},\cdots,t_{m+1})}
{\bf dt_m}
dt_{m+1}   \IEEEeqnarraynumspace
\eee}
\blankout{
\bd
\begin{array}{rcl}
\bar{\Theta}_{m,\ell}  & = &
\displaystyle \int_0^{\infty}
f_{\vec{T}_{m+1}}(t_{m+1})
\displaystyle {\bf \int}_{\bzero}^{{\bf t_{m+1}}} \\
 & \times & 
\displaystyle \frac{f_{T_1,\cdots,T_m}(t_1,\cdots,t_m)}{F_{T_1,\cdots,T_m}(t_{m+1},\cdots,t_{m+1})}
B(m,\ell,\tv)
dt_1
\cdots
dt_{m+1}
\end{array}
\ed}
\blankout{and
\bd
B(m,\ell,\tv)
\equiv
{m \choose \ell}
\prod_{j=1}^{\ell}
\bar{G}(t_{m+1} - t_j)
\prod_{k=\ell+1}^m
G(t_{m+1} - t_k) \IEEEeqnarraynumspace
\ed}
then we can express $\Hup(\Tmat)$ as
\bd
\Hup(\Tmat)
=
\sum_{\ell=1}^{M-1}
\Delta\Gamma_{M,\ell}
\log (\ell + 1)!
\ed
\end{theorem}}
\begin{Proof}{Theorem \thmref{HupGeneral}}
See the development starting in section~\ref{app:Hup} leading to the statement of \Thmref{HupGeneral}.
\end{Proof}

This concludes our calculation of $\Hup(\Tmat)$ for general input distributions $f_{\Tmat}(\cdot)$.  The
key utility of our formulation is that it does not require joint order distributions for the $\{
T_m\}$, only the more easily calculable $m^{\mbox{th}}$ order distribution for $\Tvec_m$.  We now turn to
the case where the $\Tmat$ are i.i.d. -- important because i.i.d. $\Tmat$ increases entropy
$h(\Smat)$.

\subsection{$H(\Omega|\Svv,\Tmat)$ for General IID $\Tmat$}
\label{app:HupIID}
With i.i.d. $\Tmat$, we can use the definition of $\Theta_{m,\ell}(\cdot)$ in \equat{thetadef} and the hypersymmetric result
of \equat{Hhyper} to obtain
\bee
\bar{\Theta}_{m, \ell}
=
E_{\Tvec_{m+1}} \left [
\begin{array}{c}
{m \choose \ell}
E_{T \le \Tvec_{m+1}}^{\ell} \left [
\bar{G}(\Tvec_{m+1} - T)
\right ]\\
{\mbox{\large \boldmath $\times$} }\\
E_{T \le \Tvec_{m+1}}^{m-\ell} \left [
(1 - \bar{G}(\Tvec_{m+1} - T))
\right ]
\end{array}
\right ] \IEEEeqnarraynumspace
\label{eq:thetabariid}
\eee
From the definition of $F_{\Tvec{m+1}}(\cdot)$ in \equat{ftvm} we obtain
\bee
f_{{\Tvec}_{m+1}}(t)
=
\frac{d}{dt}
\left [
1 
-
\sum_{k=0}^{m}
{M \choose k}
F_T^k(t)(1-F_T(t))^{M-k}
\right ] \IEEEeqnarraynumspace
\eee
which after rearranging as a telescoping sum simplifies to
\bee
\sum_{k=0}^{m}
(M-k)
{M \choose k}
f_T(t)F_T^k(t)(1-F_T(t))^{M-k-1} \nonumber\\
- \displaystyle \sum_{k=0}^{m-1}
(k+1)
{M \choose {k+1}}
f_T(t)F_T^{k}(t)(1-F_T(t))^{M-k-1} \IEEEeqnarraynumspace
\eee
which further simplifies to
\bee
(m+1)
{M \choose {m+1}}
f_T(t)F_T^m(t)(1-F_T(t))^{M-m-1} \IEEEeqnarraynumspace
\eee
We then define
\bee
\label{eq:phidef}
\phi(t)
=
\int_0^t
f_T(x) \bar{G}(t-x)
dx
\eee
and
\bd
\int_0^t
f_T(x) (1- \bar{G}(t-x))
dx
=
F_T(t) - \phi(t)
\ed
which allows us to write
\bd
\phi(\Tvec_{m+1}) = E_{T \le \Tvec_{m+1}} \left [
\bar{G}(\Tvec_{m+1} - T)
\right ]
\ed
and
\bd
F_T(t) - \phi(\Tvec_{m+1}) = E_{T \le \Tvec_{m+1}} \left [
1 - \bar{G}(\Tvec_{m+1} - T)
\right ]
\ed
which upon substitution into \equat{thetabariid} allows us to write $\bar{\Theta}_{m,\ell}$ as
\bea
\IEEEeqnarraymulticol{3}{l}{\bar{\Theta}_{m,\ell}   =   (m+1) {M \choose {m+1}}
{m \choose {\ell}} } \nonumber \\ \quad
 & \times  & 
\displaystyle \int_0^{\infty}
\left [
\begin{array}{c}
f_T(t)(1-F_T(t))^{M-m-1}
\phi^{\ell} (t)\\
{\mbox{\large \boldmath $\times$} }\\
(F_T(t)-\phi(t))^{m-\ell}
\end{array}
\right ]
dt
\eea
and then as
\bea
\IEEEeqnarraymulticol{3}{l}{\bar{\Theta}_{m,\ell}   = M
{{M -1} \choose \ell}
{{M - \ell - 1} \choose {m- \ell}}  
 } \nonumber \\ \quad
 & \times  & 
\begin{array}{c}
\displaystyle \int_0^{\infty}
\left [
\begin{array}{c}
f_T(t)(1-F_T(t))^{M-m-1}
\phi^{\ell} (t)\\
{\mbox{\large \boldmath $\times$} }\\
(F_T(t)-\phi(t))^{m-\ell} 
\end{array}
\right ]
dt 
\end{array}  \IEEEeqnarraynumspace
\label{eq:thetaml}
\eea
To evaluate $\Hup(\Tmat)$ in \equat{He2} we must first compute $\Gamma_{M,\ell} =\sum_{m=\ell}^{M-1}
\bar{\Theta}_{m,\ell}$ as
\bea
\IEEEeqnarraymulticol{3}{l}{\Gamma_{M,\ell} =
M {{M -1} \choose \ell}  }  \nonumber \\ \quad
 & \times & 
\int_0^{\infty}
f_T(t)
\left [
\begin{array}{c}
(1- F_T(t))^{M-1}
\left ( \frac{\phi(t)}{F_T(t) - \phi(t)} \right )^{\ell}\\
{\mbox{\large \boldmath $\times$} }\\
{\displaystyle \sum_{m=\ell}^{M-1}}
{{M - \ell - 1} \choose {m- \ell}}
\left ( \frac{F_T(t)-\phi(t)}{1 - F_T(t)} \right )^{m}
\end{array}
\right ]
dt
\IEEEeqnarraynumspace
\eee
which we rewrite as
\bea
\IEEEeqnarraymulticol{3}{l}{\Gamma_{M,\ell} =
M {{M \!-\!1} \choose \ell}  }  \nonumber \\ \quad
 & \times & 
\int_0^{\infty}
\! \!f_T(t)
\left [
\begin{array}{c}
(1\!-\! F_T(t))^{M\!-\!1}
\left ( \frac{\phi(t)}{F_T(t) - \phi(t)} \right )^{\ell}\\
{\mbox{\large \boldmath $\times$} }\\
{\displaystyle\sum_{m=0}^{M\!-\!1\!-\!\ell}}
{{M \!- \!\ell \!- \!1} \choose {m}}
\left ( \frac{F_T(t)-\phi(t)}{1 - F_T(t)} \right )^{m\!+\!\ell}
\end{array}
\right ]
dt
\IEEEeqnarraynumspace
\eee
\blankout{
\bd
\begin{array}{c}
M
{{M -1} \choose \ell}\\
{\mbox{\large \boldmath $\times$} }\\
\displaystyle \int_0^{\infty}
f_T(t)
\left [
\begin{array}{c}
(1- F_T(t))^{M-1}
\left ( \frac{\phi(t)}{F_T(t) - \phi(t)} \right )^{\ell}\\
{\mbox{\large \boldmath $\times$} }\\
{\displaystyle\sum_{m=0}^{M\!-\!1\!-\!\ell}}
{{M \!- \!\ell \!- \!1} \choose {m}}
\left ( \frac{F_T(t)-\phi(t)}{1 - F_T(t)} \right )^{m+\ell}
\end{array}
\right ]
dt
\end{array}
\ed}
We consolidate the binomial sum to obtain
\bea
\IEEEeqnarraymulticol{3}{l}{\Gamma_{M,\ell} =
M {{M \!-\!1} \choose \ell}  }  \nonumber \\ \quad
 & \times &  \int_0^{\infty}
\!\!f_T(t)
\left [
\begin{array}{c}
(1\!-\! F_T(t))^{M\!-1}
\left ( \frac{\phi(t)}{F_T(t) \!-\! \phi(t)} \right )^{\ell}\\
{\mbox{\large \boldmath $\times$} }\\
\left ( \frac{F_T(t)-\phi(t)}{1\! -\! F_T(t)} \right )^{\ell}
\left ( \frac{1-\phi(t)}{1 \!-\! F_T(t)} \right )^{M\!-1\!-\!\ell}
\end{array}
\right ]
dt   \IEEEeqnarraynumspace
\eea
which reduces to
\bee
\label{eq:gammaML}
\Gamma_{M,\ell}
=
\int_0^{\infty}
\!\!M
{{M\!-\!1} \choose \ell}
f_T(t)
\phi^{\ell}(t)
\left ( 1 \!- \!\phi(t) \right )^{M\!-1\!-\ell}
dt    \IEEEeqnarraynumspace
\eee
for $\ell = 1,2,\cdots,M-1$.

Now consider the integrand of the difference $\Gamma_{M,\ell} -
\Gamma_{M,\ell+1}$ where we drop the $t$ dependence for notational convenience
\bd
\Gamma_{M,\ell} -
\Gamma_{M,\ell+1}
=
\left [ 
\begin{array}{c}
M
{{M-1} \choose \ell}
\phi^{\ell}
\left ( 1 -\phi \right )^{M-\ell-1}\\
{\mbox{\large \boldmath $-$} }\\
M
{{M\-\!1} \choose {\ell\!+\!1}}
\phi^{\ell+1}
\left ( 1 \!-\!\phi \right )^{M\!-\!\ell-2}
\end{array}
\right ]
\ed
We can rewrite this expression as
\bd
M
\phi^{\ell}
\left [
{{M\!-\!1} \choose \ell}
\!+\!
\sum_{r=1}^{M\!-\!\ell\! -\!1}
\!\!(-1)^r
\phi^r
\left [
\begin{array}{c}
{{M-1} \choose \ell}
{{M\!-\!\ell\! -\! 1} \choose r} \\
{\mbox{\large \boldmath $+$} }\\
{{M\!-\!1} \choose {\ell\!+\!1}}
{{M\!-\!\ell\! - \!2} \choose {r\!-\!1}}
\end{array}
\right ]
\right ]
\ed
which after consolidating terms becomes
\bd
M
\phi^{\ell}
\left [
\begin{array}{c}
{{M-1} \choose \ell}\\
\mbox{ }\\
{\mbox{\large \boldmath $+$}} \\
\frac{1}{M}
{{M} \choose {\ell+1}}
{\displaystyle \sum_{r=1}^{M-\ell -1}}
(-1)^r
{{M-\ell -1} \choose {r}}
(\ell+r+1)
\phi^r
\end{array}
\right ]
\ed
Extending the sum to $r=0$ and subtracting the $r=0$ term produces
\bd
{{M} \choose {\ell+1}}
\sum_{r=0}^{M-\ell -1}
(-1)^r
{{M-\ell -1} \choose {r}}
(\ell+r+1)
\phi^{r+\ell}
\ed
which can be recognized as
\bd
\frac{d}{d\phi}
\left [
{{M} \choose {\ell+1}}
\sum_{r=0}^{M-\ell -1}
(-1)^r
{{M-\ell -1} \choose {r}}
\phi^{r+\ell+1}
\right ]
\ed
and then reduced to
\bd
{{M} \choose {\ell+1}}
\frac{d}{d\phi}
\left [
\phi^{\ell+1} (1-\phi)^{M-\ell - 1}
\right ]
\ed
so that we have $\Delta\Gamma_{M,\ell}$ as
\bee
\label{eq:Gammadiffiid}
{{M} \choose {\ell\!+\!1}}
\sum_{r=0}^{M\!-\!\ell\! -\!1}
\!\!(-1)^r
{{M\!-\!\ell\! -\!1} \choose {r}}
(\ell\!+\!r\!+\!1)
E \left [
\phi^{r\!+\!\ell}(t)
\right ]    \IEEEeqnarraynumspace
\eee
where $E[\cdot]$ is the expectation using $f_T(t)$.

The previous development of section~\ref{app:HupIID} proves the following theorem:
{\em \begin{theorem} {\bf \boldmath An Upper Bound for Ordering Entropy $H(\Omega|\Svv,\Tmat)$ with I.I.D. $\Tmat$:}
\thmlabel{DeltaGamma}

If $\Tmat$ is i.i.d., then we can write $\Delta\Gamma_{M,\ell}$ as
\bd
{{M} \choose {\ell\!+\!1}}
\sum_{r=0}^{M\!-\!\ell\! -\!1}
\!\!(-1)^r
{{M\!-\!\ell\! -\!1} \choose {r}}
(\ell\!+\!r\!+\!1)
E \left [
\phi^{r\!+\!\ell}(t)
\right ]
\ed
where 
\bd
\phi(t)
=
\int_0^t
f_T(x) \bar{G}(t-x)
dx
\ed
so that
\bd
H(\Omega|\Svv,\Tmat)
\le
\Hup(\Tmat)
=
\sum_{\ell=1}^{M-1}
\Delta\Gamma_{M,\ell}
\log (\ell + 1)!
\ed
\end{theorem}}
\begin{Proof}{Theorem \thmref{DeltaGamma}}
See the development of section~\ref{app:HupIID} leading to the statement of \Thmref{DeltaGamma}.
\end{Proof}

\subsection{$H(\Omega|\Svv,\Tmat)$  Special Case IID $\Tmat$}
\label{sect:specialcaseHO}
Here we derive expressions for $H(\Omega|\Svv,\Tmat)$ when the i.i.d. input distribution is that which maximizes
$I(\Smat;\Tmat)$.  We consider the following cases:
\begin{itemize}
\item
Exponential first-passage with $E[T] = \tau$
\item
Exponential first-passage with emission deadline, $\tau$
\end{itemize}

\subsubsection{Exponential Transit Times with a Mean Constraint}
For exponential first-passage times with mean $1/\mymu$, the probability density of
$\Tmat$ that maximizes $h(\Smat)$ subject to a mean constraint $E[\sum_m T_m]
\le M\tau$ is i.i.d. with marginal
\bee
\label{eq:fThSoptmean}
f_{T_m}(t)
=
a\delta(t)
+
\mymu a(1-a)
e^{-{\mymu}a t}
u(t)
\eee
where
$a = 1/(\mymu \tau + 1)$ \cite{bits-Qs} and $u(t)$ is the unit step function \cite{bits-Qs}.  For
exponential transit we have
\bd
\bar{G}(t)
=
e^{-\mymu t} u(t)
\ed
\blankout{
\bd
1- F_T(t) = (1-a)e^{- a \mymu t}
\ed
}
and thereby
\bd
\phi(t)
=
\int_0^t
f_T(x) \bar{G}(t-x) dx
=
a e^{-\mymu a t} u(t)
\ed
We then require an expression for $E_T[\phi^k(T)]$.
Remembering that $\int_{0^-}^{0^+} \delta (t) u^k(t) dt = \frac{1}{k+1}$ we
obtain
\bd
E_T[\phi^k(T)]
=
\int_0^{\infty} f_T(t) a^k e^{-k\mymu a t} u^k(t) dt
=
\frac{a^k}{k+1}
\ed
so that \equat{Gammadiffiid} becomes
\bd
\Delta\Gamma_{M,\ell} 
=
{{M} \choose {\ell+1}}
\sum_{r=0}^{M-\ell -1}
(-1)^r
{{M-\ell -1} \choose {r}}
a^{r+\ell}
\ed
which reduces to 
\bee
\label{eq:Gammadiffexp}
\Delta\Gamma_{M,\ell} 
=
{M \choose \ell+1}
a^\ell
(1-a)^{M-\ell - 1}
\eee
for $\ell = 1,2,\cdots,M-1$.  

With $a = \frac{1}{\mymu \tau + 1}$ we can write $H(\Omega|\Svv,\Tmat)$ as
\bee
\label{eq:HOexpgeomform}
(\mymu \tau + 1)
\sum_{k=0}^M
\log(k!)
{M \choose k}
\left (\frac{\mymu \tau}{\mymu \tau + 1} \right )^{M-k}
\left (\frac{1}{\mymu \tau + 1} \right )^k    \IEEEeqnarraynumspace
\eee
which is the expectation of $(\mymu \tau + 1) \log K!$ for a binomial random
variable $K$ with parameters $M$ and $\frac{1}{\mymu \tau + 1}$, or
\bee
\label{eq:HOexpgeomformA}
H(\Omega|\Svv,\Tmat)
=
(\mymu \tau + 1)
E_K \left [
\log K!
\right ]
\eee

We restate this result as a theorem:
{\em \begin{theorem}{\bf \boldmath $H(\Omega|\Svv,\Tmat)$ for Exponential First-Passage with a Mean Constraint ($E[T] = \tau$):}
\thmlabel{HOmaxmean}

For $T$ distributed as \equat{fThSoptmean} and exponential first-passage with parameter $\mymu$,  we have
\bd
H(\Omega|\Svv,\Tmat)
=
(\mymu \tau + 1)
E_K \left [
\log K!
\right ]
\ed
where $K$ is a binomial random variable with parameters $M$ and $\frac{1}{1+\mymu \tau}$.
\end{theorem}}
\begin{Proof}{Theorem \thmref{HOmaxmean}}
See the development leading to the statement of \Thmref{HOmaxmean} and direct application of
\Thmref{HupGeneral}.
\end{Proof}

\subsubsection{Exponential Transit Times with a Deadline}
\Thmref{fTopthSopt} states that if $T$ is constrained to $[0,\tau]$ then the $f_T(t)$ that maximizes $h(S)$
(and therefore $h(\Smat)$ when in i.i.d. form) is
\bee
\label{eq:fThSoptdeadline}
f_{T}(t)
=
{\frac{1}{e + \mymu \tau}
\delta(t)
+
\frac{\mymu}{e + \mymu \tau}
+
\frac{e-1}{e + \mymu \tau}
\delta(t-\tau)}
\eee
for $t \in [0,\tau]$ and zero otherwise.

To obtain the corresponding $H(\Omega|\Svv,\Tmat) = \Hup(\Tmat)$ we calculate $\phi(t)$ as
\bee
\label{eq:phiISalmostfs}
\int_0^{t}
f_T(x) e^{-\mymu (t-x)} dx
=
\threedef{
{\frac{1}
{e + \mymu \tau}}}{0\le t \le \tau}
{\frac{e}{e + \mymu \tau}
e^{-\mymu (t - \mymu \tau)}}{t > \tau}{0}{\mbox{o.w.}}
\eee
Once again, we require an expression for the integral $\int_0^{\infty} f_T(t)
\phi^k(t)dt$, and again remembering that 
$\int_{0^-}^{0^+} \delta (t) u^k(t) dt = \frac{1}{k+1}$ we obtain $E_T \left [ \phi^k(T) \right ]$ as
\bd
\left (\frac{1}{e+\mymu \tau} \right )^{k+1}
{\displaystyle \int_{0^-}^{0^+} }
\delta(t) 
u^k(t)
dt \\
+ \\
\mymu\left (\frac{1}{e+\mymu \tau} \right )^{k+1}
\int_0^{\tau}
dt\\
 +\\
(e\!-\!1)\left ( \frac{1}{e\!+\!\mymu \tau} \right )^{k\!+\!1}
\!\!{\displaystyle \int_{\tau^-}^{\tau^+} }
\!\!\delta(t-\tau) 
\left ( 1\! +\! (e\!-\!1)u(t\!-\!\tau) \right )^k
dt
\ed
which reduces to
\bd
\left (\frac{1}{e+\mymu \tau} \right )^{k+1}
\left [
\frac{1}{k+1} + \mymu \tau
+
\sum_{r=0}^k
{k \choose r}
\frac{1}{r+1}
(e-1)^{r+1}
\right ]
\ed
which further reduces to
\bd
\left (\frac{1}{e+\mymu \tau} \right )^{k+1}
\left [
\frac{1}{k+1} +  \mymu \tau
+
\frac{e^{k+1}}{k+1}
-
\frac{1}{k+1}
\right]
\ed
and then finally,
\bd
E_T \left [ \phi^k(T) \right ]
=
\left (\frac{1}{e+\mymu \tau} \right )^{k+1}
\left [
\mymu \tau
+
\frac{e^{k+1}}{k+1}
\right]
\ed
so that $\Delta\Gamma_{M,\ell}$ in \equat{Gammadiffiid} becomes
\bd
{{M} \choose {\ell\!+\!1}}
\displaystyle \sum_{r=0}^{M\!-\!\ell\! -\!1}
\left [
\begin{array}{c}
(-1)^r
{{M-\ell -1} \choose {r}}\\
\times \\
\!\!(\ell\!+\!r\!+\!1)
\left (\frac{1}{e+\mymu \tau} \right )^{r\!+\!\ell\!+\!1}
\!\!\left [
\mymu \tau
\!+\!
\frac{e^{r+\ell+1}}{r+\ell+1}
\right]
\end{array}
\right ]
\ed
which reduces to 
\bd
{{M} \choose {\ell+1}}
\left (\frac{e}{e+\mymu \tau} \right )^{\ell+1}
\left (\frac{\mymu \tau}{e+\mymu \tau} \right )^{M-\ell-1}\\
{\mbox{\large \boldmath $+$}}\\
\mymu \tau
{{M} \choose {\ell+1}}
\displaystyle \sum_{r=0}^{M-\ell -1}
\left [
\begin{array}{c}
(-1)^r
{{M-\ell -1} \choose {r}}\\
\times\\
(\ell+r+1)
\left (\frac{1}{e+\mymu \tau} \right )^{r+\ell+1}
\end{array}
\right ]
\ed
and then to
\bd
{{M} \choose {\ell+1}}
\left (\frac{e}{e+\mymu \tau} \right )^{\ell+1}
\left (\frac{\mymu \tau}{e+\mymu \tau} \right )^{M-\ell-1}\\
{\mbox{\large \boldmath $+$}}\\
\left [
\begin{array}{c}
\mymu \tau
{{M} \choose {\ell+1}}
\left (1 - \frac{1}{e+\mymu \tau} \right )^{M-\ell -2}\\
\times \\
\left (\frac{1}{e+\mymu \tau} \right )^{\ell +1}
\left ( \ell + 1 -\frac{M}{e+\mymu \tau} \right )
\end{array}
\right ]
\ed
If we define $k = \ell +1$ and then
\bd
p_1 = \frac{e}{e+\mymu \tau}
\ed
and
\bd
p_2 = \frac{1}{e+\mymu \tau}
\ed
we can then write 
\bee
\label{eq:Gammadiffuniform}
\Delta\Gamma_{M,k-1}
=
\left [
\begin{array}{c}
{M \choose k}
p_1^k (1-p_1)^{M-k}\\
{\mbox{\large \boldmath $+$}}\\
\frac{\mymu \tau }{1-p_2}
\left [
k
-
\frac{M}{\mymu \tau + e}
\right ]
{M \choose k}
p_2^k (1-p_2)^{M-k}
\end{array}
\right ] \IEEEeqnarraynumspace
\eee
Now if we define random variables $K_i$ to be binomial with parameters $M$ and
$p_i$, the following theorem results from direct application of \Thmref{HupGeneral}:
{\em \begin{theorem}{\bf \boldmath $H(\Omega|\Svv,\Tmat)$ for Exponential First-Passage with a Launch Deadline ($\Tmat \in [0,\tau]^M$):}
\thmlabel{HOmaxdeadline}
      
For $T$ distributed as \equat{fThSoptdeadline} we have
\bee
\label{eq:HOexpgeomform2}
H(\Omega|\Svv,\Tmat)
=
\!\left [
\!\!\begin{array}{c}
E_{K_1} \left [
\log K_1!
\right ]
\!+\!
\frac{\mymu \tau}{1-p_2}
E_{K_2} \left [
K_2 \log K_2!
\right ]\\
{\mbox{\large \boldmath $-$}}\\
\frac{\mymu \tau M }{(1-p_2)(\mymu \tau + e)}
E_{K_2} \left [
\log K_2!
\right ]
\end{array}
\! \!\right ] \IEEEeqnarraynumspace
\eee
where $K_1$ is a binomial random variable with parameters $M$ and $\frac{e}{e+\mymu \tau}$
and $K_2$ is a binomial random variable with parameters $M$ and $\frac{1}{e+\mymu \tau}$.
\end{theorem}}
\begin{Proof}{Theorem \thmref{HOmaxdeadline}}
See the development leading to the statement of \Thmref{HOmaxdeadline} and direct application of
\Thmref{HupGeneral}.
\end{Proof}
\subsection{Asymptotic $H(\Omega|\Svv,\Tmat)/M$ For Exponential First-Passage}
We are interested in asymptotic values of $H(\Omega|\Svv,\Tmat)/M$ owing to our definition of
capacity per token in \equat{Cqdef} (see also in Part-I \cite{RoseMian16_1}).  To that end, recall
that $\mylambda \tau = M$ and we define $\myrho = \mylambda/\mymu$, a measure of system token
``load'' (also a proxy for power expenditure in units of energy per passage time), so that
\bd
\frac{1}{1 +  \mymu M/\mylambda}
=
\frac{1}{1 +  M/\myrho}
\ed
and likewise
\bd
\frac{e}{e +  \mymu M/\mylambda}
=
\frac{e}{e + M/\myrho}
\ed
and
\bd
\frac{1}{e +  \mymu M/\mylambda}
=
\frac{1}{e + M/\myrho}
\ed
Now, remember the binomial distribution for fixed $k$ and large $M$ is approximated by
\bd
{M \choose k} p^k (1-p)^{M-k}
\approx
\frac{M^k}{k!}
p^k (1-p)^{M-k}
\ed
So, for any finite $k$ it is easily seen that for $M \rightarrow \infty$
\bee
\label{eq:limit1}
{M \choose k} 
\left ( \frac{1}{1 + \frac{M}{\myrho}} \right )^k
\left (1 -  \frac{1}{1 + \frac{M}{\myrho}} \right )^{M-k}
\rightarrow
e^{-\myrho}
\frac{1}{k!}
\myrho^k  \IEEEeqnarraynumspace
\eee
\bee
\label{eq:limit2}
{M \choose k} 
\left ( \frac{1}{e + \frac{M}{\myrho}} \right )^k
\left (1 -  \frac{1}{e + \frac{M}{\myrho}} \right )^{M-k}
\rightarrow
e^{-\myrho}
\frac{1}{k!}
\myrho^k  \IEEEeqnarraynumspace
\eee
and
\bee
\label{eq:limit3}
{M \choose k} 
\left ( \frac{e}{e + \frac{M}{\myrho}} \right )^k
\left (1 -  \frac{e}{e + \frac{M}{\myrho}} \right )^{M-k}
\rightarrow
e^{-\myrho e}
\frac{1}{k!}
\myrho^k e^k   \IEEEeqnarraynumspace
\eee
and we note that all these limiting distributions are Poisson.

\Equat{HOexpgeomformA} and \equat{HOexpgeomform2} can then be combined with \equat{limit1},
\equat{limit2} and \equat{limit3} to produce the following two theorems:
{\em \begin{theorem}{\bf \boldmath Asymptotic $H(\Omega|\Svv,\Tmat)/M$ for Exponential First-Passage with a Mean Constraint ($E[T] = \tau$):}
\thmlabel{HOasymptmean}

For exponential first-passage with $E[T] = \tau$ and $f_T(\cdot)$ as given in \equat{fThSoptmean}, $H(\Omega|\Svv,\Tmat)$ is given by 
\bee
\label{eq:HOiidmean}
\lim_{M \rightarrow \infty} \frac{H(\Omega|\Svv,\Tmat)}{M}
=
e^{-\myrho}
\sum_{k=2}^{\infty}
\myrho^{k-1}
\frac{\log k!}{k!}
=
E[\frac{\log k!}{\rho}]   \IEEEeqnarraynumspace
\eee
where the final expectation is for $k$ a Poisson random variable with parameter $\rho$.
\end{theorem}}
\begin{Proof}{Theorem \thmref{HOasymptmean}}
See \Thmref{HOmaxmean} and the development leading up to the statement of \Thmref{HOasymptmean}.
\end{Proof}

{\em \begin{theorem}{\bf \boldmath Asymptotic $H(\Omega|\Svv,\Tmat)/M$ for Exponential First-Passage with a Deadline Constraint ($T \in [0,\tau]$)}
\thmlabel{HOasymptdeadline}

For exponential first-passage with $\Tmat \in [0,M/\myrho]^M$ and $f_T(\cdot)$ as given in \equat{fThSoptdeadline}, 
$H(\Omega|\Svv,\Tmat)$ is given by 
\bee
\label{eq:HOiiddeadline}
\lim_{M \rightarrow \infty} \frac{H(\Omega|\Svv,\Tmat)}{M}
=
E[ (\frac{k}{\rho} - 1) \log k!]  \IEEEeqnarraynumspace
\eee
where the final expectation is for $k$ a Poisson random variable with parameter $\rho$.
\end{theorem}}
\begin{Proof}{Theorem \thmref{HOasymptdeadline}}
See \Thmref{HOmaxdeadline} and the development leading up to the statement of \Thmref{HOasymptmean}.
\end{Proof}

\blankout{
\section{Lower Bound on $C_q$}
{\em \begin{theorem} {\bf \boldmath A Lower Bound on $C_q$:}
\thmlabel{Ilowbound}

If the first-passage density $f_D(\cdot)$ is exponential with parameter $\mymu$ and $\mylambda$ is the constant
average rate at which tokens are released ($\mylambda \equiv \lim_{M \rightarrow \infty}\frac{M}{\tau(M)}$) then the capacity \cite{isit13} per token, $C_q$, obeys
\bee
\label{eq:Cmdown}
C_q
\ge
-\log \myrho + 
e^{-\myrho}
{\displaystyle \sum_{k=1}^{\infty}}
{\myrho}^k
(\frac{k}{\myrho} - 1)
\frac{\log k!}{k!}
\eee
where $\myrho \equiv \frac{\mylambda}{\mymu}$.
\end{theorem}}
\begin{Proof}{Theorem \thmref{Ilowbound}}
See development leading to \Thmref{Ilowbound}.
\end{Proof}

We can rewrite \equat{Cmdown} more compactly by noting that
\bas
\begin{array}{rcl}
{\displaystyle \sum_{k=1}^{\infty}}
\myrho^k
(\frac{k}{\myrho} - 1)
\frac{\log k!}{k!} & = & {\displaystyle \sum_{\ell=1}^{\infty}}
\log \ell
{\displaystyle \sum_{k=\ell}^{\infty}}
\myrho^k
\frac{(\frac{k}{\myrho} - 1)}{k!}
\end{array}
\eas
Then
\bd
\sum_{k=\ell}^{\infty}
\left (\myrho \right )^k \frac{1}{k!}
=
e^{\myrho}
-
\sum_{k=0}^{\ell - 1}
\left (\myrho \right )^k \frac{1}{k!}
\ed
and
\bd
\sum_{k=\ell}^{\infty} 
k/ \myrho \left (\myrho \right )^k \frac{1}{k!}
=
\sum_{k=\ell-1}^{\infty} 
\left (\myrho \right )^k \frac{1}{k!}
\ed
\blankout{
which leads to
\bd
\sum_{k=\ell}^{\infty}
(k/\myrho - 1)\left  (\myrho \right )^k \frac{1}{k!} 
=
\sum_{k=\ell-1}^{\infty} 
\left (\myrho \right )^k \frac{1}{k!}
+
\sum_{k=0}^{\ell - 1}
\left (\myrho \right )^k \frac{1}{k!}
-
e^{\myrho}
\ed}
can be used to obtain
\bd
\sum_{k=\ell}^{\infty}
\left  (\myrho \right )^k \frac{(k/\myrho - 1)}{k!} 
=
\frac{1}{(\ell - 1)!} \left  (\myrho \right )^{\ell -1}  
=
\ell \left  (\myrho \right )^{\ell}  \frac{1}{\myrho \ell!} 
\ed
We can then define the probability mass function
$
p_\ell
=
e^{-\myrho}
\left (\myrho \right )^\ell \frac{1}{\ell!}
$, $\ell = 0, 1, \cdots, \infty$ to obtain the more compact
\bee
\label{eq:HOsimple}
{\displaystyle \sum_{k=1}^{\infty}}
\left (\myrho \right )^k
(k/ \myrho - 1)
\frac{\log k!}{k!}
=
\frac{1}{\myrho} E_\ell \left [ (\ell \log \ell) \right ]
\eee
so that \equat{Cmdown} becomes
\bee
\label{eq:Cmdownsimple}
C_m
\ge
-\log \myrho + 
\frac{1}{\myrho} E_\ell \left [ (\ell \log \ell) \right ]
\eee
}

\blankout{
\section{Simple Lower Bound for $C_q$}
We can derive a simple lower bound on $C_q(M)$ by noting that
\equat{orderedMI_decomp} and the definition of \equat{pertokenCtauM} with
$\tau(M)$ produces
\bee
\label{eq:RtauMboundsimple}
\begin{array}{rcl}
C_q(M) & = & 
\max_{f_{\Tmat}(\cdot)}
\left [  I(\Smat;\Tmat) + H(\Omega|\Svv,\Tmat) \right ]
-
\log M!\\
 & \ge & \max_{f_{\Tmat}(\cdot)}
I(\Smat;\Tmat) - \log M!
\end{array}
\eee
because $0 \le H(\Omega|\Svv,\Tmat) \le M!$.

From section~\ref{sect:minmaxhS} (and \cite{isit11}) we know that the univariate maximum $I(S;T)$
subject to $T \le \tau$ and a mean first-passage time $\mymu^{-1}$ is also minimized when the mean
first-passage time density $g(\cdot)$ is exponential with parameter $\mymu$.  We therefore have for any
finite $M$ and a finite launch deadline $\tau(M)$,
\bee
\label{eq:Iminmaxdeadline}
\max_{f_{\Tmat}(\cdot)}
I(\Smat;\Tmat)
\ge
\min_{g(\cdot)} \max_{f_{\Tmat}(\cdot)} I(\Smat;\Tmat)
=
M \log \left (1 + \frac{\mylambda \tau(M)}{e} \right )
\eee
which means,
\bee
C_q(M)
\ge
\log \left (1 + \frac{\mylambda \tau(M)}{e} \right ) - \frac{\log(M!)}{M}
\eee
for a launch deadline $\tau(M)$.

Using $\myrho = \mylambda/\mymu$ and Stirling's approximation, $\log M! = M \log M - M +
O(\log(M))$ we have the following sequence of simplifications
\bd
\begin{array}{c}
\frac{1}{M} \left ( M \log \left ( 1 +  \frac {\mylambda}{\mylambda e} M \right ) - \log M! \right )\\
\log \left ( 1 + \frac{\mylambda}{\mylambda e} M \right )
- \log M + 1 - \frac{1}{M}O(\log(M))\\
\log \left ( \frac{e}{M}  + \frac{\mylambda}{\mylambda} \right )
- \frac{1}{M}O(\log(M))
\end{array}
\ed
Defining $\myrho = \frac{\mylambda}{\mymu}$, the ratio of the token uptake rate to the
release rate, and then taking the limit as $M \rightarrow \infty$ we obtain
\bee
\lim_{M \rightarrow \infty}
C_q(M)
\ge
-\log \myrho
\eee
The above results prove the following theorem:
{\em \begin{theorem} {\bf \boldmath A Simple Lower Bound for $C_q$:}
  \thmlabel{IDlowerbound}

Given an average rate of signaling token production $\mylambda = M/\tau$
and any i.i.d. first-passage time distribution with mean $\mymu^{-1}$,
the timing channel capacity $C_q(\rho)$ in nats per token obeys
\bee
\label{eq:Cqlo}
C_{q}(\myrho) \ge \max \left
    \{-\log\myrho,0 \right \}
\eee
where $\myrho = \frac{\mylambda}{\mymu}$
\end{theorem}}
\begin{Proof}{Theorem \thmref{IDlowerbound}}
See development leading to \Thmref{IDlowerbound}.
\end{Proof}

We emphasize that the Theorem \thmref{IDlowerbound} bound is {\em general} and
applies to {\em any} first-passage time density $g(\cdot)$ with mean $\mymu^{-1}$.

}

\section{Upper Bound for $I(\Svv;\Tmat)$}
\label{sect:upperI}
With analytic bounds for $H(\Omega|\Svv,\Tmat)$, we can now consider bounds on the mutual information,
$I(\Svv;\Tmat)$.  In Part-I (using results from this, Part-II) lower bounds were derived.  Here we consider an upper bound. To begin, however, we must find an upper bound for $H(\Omega|\Svv,\Tmat)$.

\subsection{A Useful Upper Bound On $H(\Omega|\Svv,\Tmat$)}
\label{sect:upperHO}
We state the bound as a theorem with proof.
{\em \begin{theorem} {\bf \boldmath An Upper Bound for $H(\Omega|\Svv,\Tmat)$:}
\thmlabel{HOgamma}

Given
\bee
Q(\cdot) = \bG(|\cdot|)
\eee
where $\bG(\cdot)$ is the CCDF of the passage time, and defining
\bee
\label{eq:gamma}
\gamma_T = E_{\Tmat} \left [ Q(T_1  - T_2) \right ] 
\eee
we have
\bee
\label{eq:Homega_bound}
H(\Omega|\Svv,\Tmat)
\le 
E_{\Tmat}\left [ \Hup(\Tmat) \right ]
\le
M \log \left ( 1 + \frac{M-1}{2} \gamma_T \right )  \IEEEeqnarraynumspace
\eee
\end{theorem}}
\begin{Proof}{Theorem \thmref{HOgamma}}
$\Hup(\tv)$, defined in \equat{Hexpurg3} and derived in Part-I \cite{RoseMian16_1, SonRos13} is an
  upper bound for $H(\Omega|\Svv,\tvv)$.  The bound is satisfied with equality {\bf iff} the
  first-passage density is exponential \cite{RoseMian16_1,SonRos13}.  For a given $m$, let us define
  $\bG_k = \bG(\tvec_{m+1} - \tvec_k)$ and $G_k$ in a corresponding way.  Then, consider the sum of
  the following $2^m$ terms
\bd
\bG_{m} \bG_{m-1} \bG_{m-2}  \cdots \bG_3 \bG_2 \bG_1\\
+\\
\bG_{m} \bG_{m-1} \bG_{m-2}  \cdots \bG_3 \bG_2  G_1\\
+\\
\vdots\\
+\\
G_{m} G_{m-1} G_{m-2}  \cdots   G_3 G_2 \bG_1\\
+\\
G_{m} G_{m-1} G_{m-2}  \cdots   G_3 G_2 G_1 
\ed
Taken pairwise it is easy to see that this sum telescopes to $1$ since $\bG_i + G_i = 1$ so that the
ensemble of terms is a PMF.  Furthermore, since $m = 1,2, \cdots, M$, the complete ensemble of the
terms, $\prod_{j=1}^m \bG^{\bar{x}_j}(\tvec_{m+1} - \tvec_j) G^{1-\bar{x}_j}(\tvec_{m+1} -
\tvec_j)$, $m=1,2,\cdots, M$, sums to $M$. So, we can define
\bee
\label{eq:pellm}
p_{\ell|\tvv,m}
= 
\sum_{|\bar{\xv}| = \ell}
\prod_{j=1}^m
{\bG}^{\bar{x}_j}(\tvec_{m+1} - \tvec_j)
G^{1 -\bar{x}_j}(\tvec_{m+1} - \tvec_j)  \IEEEeqnarraynumspace
\eee
and then
\bee
\label{eq:pell}
p_{\ell|\tvv}
= 
\sum_{m=\ell}^{M-1}
\sum_{|\bar{\xv}| = \ell}
\prod_{j=1}^m
\frac{{\bG}^{\bar{x}_j}(\tvec_{m+1} - \tvec_j)
G^{1 -\bar{x}_j}(\tvec_{m+1} - \tvec_j)}{M}  \IEEEeqnarraynumspace
\eee
for $\ell = 0, 1, \cdots, M-1$. We can use Jensen's inequality to write
\bee
\label{eq:hup}
\Hup(\tv) = 
E_{\ell|\tvv} \left [\log(1+\ell)  \right ]
\le  M \log (E[\ell|\tvv] + 1)
\eee
Now consider that 
\bd
E[\ell|\tvv]
=
\sum_{m=0}^{M-1}
\frac{1}{M}
E[\ell|\tvv,m]
\ed
and the explicit expansion of $E[\ell|\tvv,m]$ is
\bee
\label{eq:El}
\sum_{\ell=0}^m \ell
\mbox{\small $
\left (
{\displaystyle \sum_{|\bar{\xv}| = \ell}
\prod_{j=1}^m}
{{\bG}^{\bar{x}_j}(\tvec_{m+1} - \tvec_j)
{G}^{1- \bar{x}_j}(\tvec_{m+1} - \tvec_j)}
\right )$} \IEEEeqnarraynumspace
\eee

Then consider that $E[\ell| \tvv,m]$ has the terms
\bd
\label{eq:Elgroup}
\left .
0 \times
\left [
\begin{array}{c}
G_{m}  G_{m-1} G_{m-2} \cdots   G_3 G_3 G_1 \nonumber \\ 
\end{array}
\right ] \right \}  {\mbox{$1$ term}}
\ed
\bd
\left .
1 \times
\left [
\begin{array}{c}
G_{m} G_{m-1} G_{m-2}  \cdots   G_3 G_2 \bG_1\nonumber \\
+\nonumber \\
G_{m} G_{m-1} G_{m-2}  \cdots   G_3 \bG_2 G_1\nonumber \\
+\nonumber \\
\vdots\nonumber \\
+\nonumber \\
\bG_{m} G_{m-1} G_{m-2}  \cdots   G_3 G_2 G_1
\end{array}
\right ] \right \}  {\mbox{$m$ terms}}
\ed
\bd
\left .
2 \times
\left [
\begin{array}{c}
G_{m} G_{m-1} G_{m-2}  \cdots G_3 \bG_2 \bG_1\nonumber \\
+\nonumber \\
\vdots\nonumber \\
+\nonumber \\
\bG_{m} \bG_{m-1} G_{m-2}  \cdots G_3 G_2 G_1
\end{array}
\right ]  \right \}{\mbox{${m \choose 2}$ terms}}
\ed
with final term
\bd
\left .
m \times
\left [
\begin{array}{c}
\bG_{m} \bG_{m-1} \bG_{m-2} \cdots \bG_{3} \bG_{2}\bG_1 \\
\end{array}
\right ] \right \}{\mbox{$1$ term}}
\ed
Then consider the term $G_{m} G_{m-1} G_{m-2} \cdots G_2 \bG_1$ and group
together the other $2^{m-1} - 1$ different terms that contain $\bG_1$. The sum of
all these terms is $\bG_1$.  We can do a corresponding
grouping for each of the $m$ terms in which $\bG_i$ appears exactly once.

Thus, by expanding and regrouping  the inner product terms of \equat{El} we can show that
\bd
E[\ell|\tvv,m] 
=
\sum_{j=1}^m  {\bG}(\tvec_{m+1} - \tvec_j)
\ed
which results in
\bd
\Hup(\tv)
\le 
M \log \left (
1 + \frac{1}{M} \sum_{m=1}^{M-1} \sum_{j=1}^m {\bG}(\tvec_{m+1} - \tvec_j)
\right )
\ed
via \equat{pell} and \equat{hup}, remembering that $E[\ell|\tvv,m=0] = 0$. Taking the expectation in
$\Tvv$ yields
\bee
\label{eq:EHup}
\Hup(\Tmat)
\le 
M \log \left (
1 + \sum_{m=1}^{M-1} \sum_{j=1}^m 
\frac{E \left[ {\bG}(\Tvec_{m+1} - \Tvec_j) \right ]}{M}
\right ) \IEEEeqnarraynumspace
\eee
We then note that all ordered differences between the $T_i$ are
accounted for in \equat{EHup}. For any given $\Tmat$ there are $\frac{M(M-1)}{2}$ ordered
terms. Thus, we can rewrite \equat{EHup} as
\bee
\label{eq:EHup2}
E_{\Tvv}\left [ 
\Hup(\Tmat)
\right ]
\le
M \log \left (
1 +  \sum_{i,j, i \ne j}^M 
\frac{E \left[ {\bG}(\left |T_i - T_j \right |) \right ]}{2M}
\right ) \IEEEeqnarraynumspace
\eee
where the factor of $\frac{1}{2}$ is introduced to account for terms
$T_i < T_j$ which would not appear in the ordered case of \equat{EHup}.
Finally, hypersymmetry of $\Tmat$ requires that $E \left[ {\bG}(\left |T_i - T_j \right
  |) \right ] = \gamma_T$, a constant for $i\ne j$ so that
\bd
H(\Omega|\Svv,\Tmat)
\le 
E_{\Tvv}\left [ 
\Hup(\Tmat)
\right ]
\le
M \log \left (
1 + \frac{M-1}{2} \gamma_T
\right )
\ed
which matches the result stated in Theorem \thmref{HOgamma} and thus proves the theorem.
\end{Proof}

\subsection{Maximizing $h(\Smat) + M \log \left ( 1 + \gamma_S (M-1) \right )$}
We now have the rudiments of an upper bound for $I(\Svv;\Tmat)$ in
\bea
\IEEEeqnarraymulticol{3}{l}{\max_{f_{\Tmat}(\cdot)} \frac{I(\Svv;\Tmat)}{M}} \nonumber \\ \quad
 &  \le & 
\frac{h(\Smat)}{M}
+
\log \left ( 1 + \gamma_S (M-1) \right )
- \frac{\log M!}{M} - h(D) \IEEEeqnarraynumspace
\eea
However, the upper bound \equat{Homega_bound} is in terms of $f_{\Tmat}(\cdot)$ whereas $h(\Smat)$
is a function(al) of $f_{\Smat}(\cdot)$.  Therefore, we must develop a relationship between
$\gamma_T = E \left [ Q(T_1 - T_2) \right ]$ and $\gamma_S = E \left [ Q(S_1 - S_2) \right ]$. This
relationship allows us to fix $\gamma_S$ and maximize $h(\Smat)$ while still maintaining an upper
bound on $H(\Omega|\Svv,\Tvv)$.  From here onward we assume exponential first-passage of tokens.

{\em \begin{theorem} {\bf \boldmath $\gamma_T$ versus $\gamma_S$ for Exponential First-Passage:}
\thmlabel{EQTS}

If the first-passage density $f_D(\cdot)$ is exponential then
\bd
E \left [ Q(S_1  - S_2) \right ]
\ge
\frac{1}{2} E \left [ Q(T_1  - T_2) \right ] 
\ed
or
\bd
\gamma_S
\ge
\frac{1}{2}
\gamma_T
\ed
\end{theorem}}
\begin{Proof}{Theorem \thmref{EQTS}}
Let $\Delta = T_1 - T_2$ and ${\cal D} = D_2 - D_1$.  Then $\Delta+{\cal D} =
S_1 - S_2$.  For the i.i.d. $D_i$ exponential we have ${\bG}(d) = e^{-\mylambda d}$,
$d \ge 0$.  Thus, $Q(\cdot)= e^{-\mylambda |\cdot|}$.  We then note that $|a+b| \le
|a| + |b|$ so that 
\bas
E[Q(\Delta+{\cal D})]  & = & E[e^{-\mylambda | \Delta+{\cal D} |}]\\
&  \ge & E[e^{-\mylambda |\Delta| -\mylambda |{\cal D} |}]\\
 & = &  E[Q(\Delta)]E[Q({\cal D})]
\eas
because $\Delta$ and ${\cal D}$ are independent.  Then consider that the density of ${\cal D}$
is $f_{\cal D}({\cdot}) = \frac{\mylambda}{2}e^{-\mylambda |{\cdot}|}$ so that
$E[Q({\cal D})] = \int_{=\infty}^\infty \frac{\mylambda}{2}e^{-\mylambda |z|} e^{-\mylambda|z|} dz =
\frac{1}{2}$ which completes the proof.
\end{Proof}

Now, suppose we fix $E \left [ Q(S_1 - S_2) \right ] = \gamma_S$. Then, owing to hypersymmetry we have
$E \left [ Q(S_i - S_j) \right ] = \gamma_S$  $\forall i,j,i \ne j$.  Using
standard Euler-Lagrange optimization \cite{hild}, we can find the density
$f_{\Smat}$ which maximizes $h(\Smat)$ as
\bee
f_{\Smat}^* (\sv) = \frac{1}{A(\beta)} e^{\beta \sum_{\stackrel{i,j}{i \ne j}}Q(s_i - s_j)}
\eee
where
\bee
A(\beta)
=
\int e^{\beta \sum_{\stackrel{i,j}{i\ne j}}Q(s_i - s_j)} d\sv
\eee
and $\beta$ is a constant chosen to satisfy  $E[Q(S_1 - S_2)] = \gamma_S$.
The entropy  of $\Smat$ is then
\bee
h(\Smat)
=
\log A(\beta) - \beta M(M-1) \gamma_S
\eee
We note that for $\beta=0$, $f_{\Smat}(\cdot)$ is uniform.  Increasing $\beta$ makes
$f_{\Smat}(\cdot)$ more ``peaky'' in regions where $s_i \approx s_j$ since $Q(0)=1$
and $Q(\cdot)$ is monotonically decreasing away from zero. Likewise, decreasing
$\beta$ reduces $f_{\Smat}(\cdot)$ in the vicinity of $s_i \approx s_j$.  Thus,
$\gamma_S$ increases monotonically with $\beta$.   The result is that
$\gamma_S^{\prime}(\cdot)$ is strictly positive.

More formally, we have from the definition of $\gamma_S(\beta)$ that
\bd
M(M-1)\gamma_S(\beta)
=
E\left [\sum_{\stackrel{i,j}{i\ne j}}Q(s_i - s_j) \right ]
\equiv
\Gamma_S(\beta)
\ed
Then
\bee
\label{eq:variance}
\Gamma_S^{\prime}(\beta)
\!=\!
E\!\!\left [ \!\!\left ( \! \sum_{\stackrel{i,j}{i\ne j}}\!Q(s_i\! - \!s_j) \!\right )^{\!2} \right ]
\!-\!
E^2 \!\!\left [\sum_{\stackrel{i,j}{i\ne j}}\!Q(s_i\! -\! s_j) \! \right ]  \IEEEeqnarraynumspace
\eee
which is a variance and therefore greater than or equal to zero.  Thus,
$\gamma_S^{\prime}(\beta) \ge 0$.  And since $0 \le \gamma_S(\beta) \le 1$, we
must also have $\gamma_S^{\prime}(\beta) \rightarrow 0$ in the limits $\beta
\rightarrow \pm \infty$.

Now, consider all terms as functions of $\beta$ as in
\bee
\label{eq:Ibeta}
\begin{array}{rcl}
I(\Svv;\Tmat) & \le &  \log A(\beta) - \beta M(M-1)\gamma_S(\beta) \\
 & + & M \log \left ( 1 + \gamma_S(\beta)(M-1) \right )\\
 &  - &  h(\Smat|\Tmat) - \log M!
\end{array}
\eee
We can find extremal points by differentiating \equat{Ibeta} with respect to $\beta$ to obtain
the first derivative
\bd
M(M-1) \gamma^{\prime}_S (\beta) \left (
-\beta 
+
\frac{1}{1 + \gamma_S(\beta)(M-1)}
\right )
\ed
and the second derivative
\bd
\begin{array}{c}
M(M-1)\gamma^{\prime\prime}_S(\beta)
\left (
-\beta 
+
\frac{1}{1 + \gamma_S(\beta)(M-1)}
\right )\\
+\\
-M(M-1) \gamma^{\prime}_S (\beta) \left (
1 + (M-1)\frac{\gamma^{\prime}_S(\beta)}{\left ( 1 + \gamma_S(\beta)(M-1) \right )^2}
\right )
\end{array}
\ed
which when the first derivative is zero reduces to 
\bd
-M(M-1) \gamma^{\prime}_S (\beta) \left (
1 + (M-1)
\begin{array}{c}
  \frac{\gamma^{\prime}_S(\beta)}{\left ( 1 + \gamma_S(\beta)(M-1) \right )^2}
\end{array}
\right ) \le 0  \IEEEeqnarraynumspace
\ed
We then have
\bee
\label{eq:betastar}
\gamma_S^* = \gamma_S(\bstar)
=
\frac{1 - \bstar}{(M-1)\bstar}
\eee
and note that \equat{betastar} requires $\frac{1}{M} \le \bstar \le 1$ since $0
\le \gamma_S(\beta) \le 1$. In addition, there is at most one solution to
\equat{betastar} since $\frac{1 - \bstar}{(M-1)\bstar}$ monotonically
decreases in $\beta$ while $\gamma_S(\beta)$ monotonically increases in $\beta$.
Since the second derivative at the extremal is non-positive, the unique point
defined by \equat{betastar} is a maximum.

Unfortunately, solutions to \equat{betastar} have no closed form and numerical
solutions for asymptotically large $M$ are impractical.  Nonetheless, the constraints
on $\bstar$ will allow an oblique approach to deriving a bound.

We note again that $\Gamma_S^{\prime}(\beta)$, is the variance of $\sum_{i\ne j}Q(s_i
- s_i)$ and must decrease monotonically in $\beta$ since as previously
discussed, increased $\beta$ concentrates $f_{\Smat}(\cdot)$ around larger values of
$\sum_{i\ne j}Q(s_i - s_i)$.  Thus,
\bee
\Gamma_S^{\prime}(\beta) \le \Gamma_S^{\prime}(0)
\eee
$\forall \beta > 0$ which in turn implies
\bee
\label{eq:diffeq}
\Gamma_S(\beta) \le  \beta \Gamma_S^{\prime}(0)   + \Gamma_S(0)
\eee
$\forall \beta \in (0,1]$.

Assuming exponential first-passage, $Q(x) = e^{-\mu|x|}$ and remembering that
$\Gamma_S(\beta) = M(M-1)\gamma_S(\beta)$, we can calculate both $\gamma_S(0)$
and $\gamma_S^{\prime}(0)$ in closed form as
\bee
\label{eq:Gamma0}
\gamma_S(0)
=
Z(\mu \tau)
\equiv
\frac{2}{(\mu \tau)^2} \left ( \mu \tau + e^{-\mu \tau} - 1 \right )
\eee
and
\bee
\label{eq:Gammaprime0}
\gamma_S^{\prime}(0)
=
\left [
\begin{array}{c}
(M-2)(M-3) \gamma_S^2(0) + 2Z(2 \mu\tau)\\
+ \\
24\frac{M-2}{(\mu \tau)^3} \left ( \mu \tau - 2 + e^{-\mu \tau} (2 + \mu \tau) \right ) \\
-\\
M(M-1) \gamma_S^2(0)
\end{array}
\right ]
\eee
respectively.   Defining $M = \mylambda \tau$ and taking the limit for large $M$ yields
\bee
\label{eq:limgamma}
\lim_{M \rightarrow \infty} M \gamma_S(0) = \frac{2\mylambda}{\mymu} = 2\rho
\eee
and
\bee
\label{eq:limgammaprime}
\lim_{M \rightarrow \infty}
(M-1)\gamma_S^{\prime}(0)
=
8 \frac{\mylambda^2}{\mu^2}
+
2 \frac{\mylambda}{\mu}
=
8\myrho^2 + 2\myrho
\eee
where once again $\myrho = \frac{\mylambda}{\mymu}$.

Remembering that $\Gamma_S(\beta) = M(M-1)\gamma_S(\beta)$ and utilizing \equat{diffeq} we have
\bee
\label{eq:gammaprimemax}
\gamma_S(0)
\le 
\gamma_S(\bstar)
\le
\gamma_S^{\prime}(0) \bstar + \gamma_S(0)
\eee
Thus, the $\gamma$-intercept of the monotonically decreasing $\frac{1-\beta}{(M-1)\beta}$
with the right hand side of \equat{gammaprimemax} must yield a value at least as large as
$\gamma(\bstar)$.  To solve for this intercept we set
\bea
\frac{1-\tbeta}{(M-1)\tbeta} & = & \gamma_S^{\prime}(0) \tbeta + \gamma_S(0) \nonumber \\
 & = & 
\frac{1}{M-1} \tbeta \left ( {8\myrho^2} + {2\myrho} \right ) + {2\myrho}\frac{1}{M} \IEEEeqnarraynumspace
\eea
so that in the limit of large $M$ we have
\bd
\tbeta
=
\frac{\sqrt{1 + {12 \myrho} +36 {\myrho^2}} - (1 + {2\myrho})}{{16 \myrho^2} + {4\myrho}}
=
\frac{1}{4 \myrho + 1}
\ed
which results in
\bee
\label{eq:tgamma}
(M-1) \gamma(\bstar)
\le
\frac{1}{4 \myrho + 1}
\left (8 \myrho^2 +  2 \myrho \right ) +  {2\myrho}
=
4 \myrho
\eee
so that for large $M$ we have
\bea
I(\Svv;\Tmat) & \le &  \log A(\bstar)  - \bstar M  (M-1)\gamma_S(\bstar) \nonumber\\
 & + & M \log \left ( 1 +  4 \myrho \right )
 -  h(\Smat|\Tmat) - \log M! 
\label{eq:Ibetamax1}
\eea

To complete the mutual information bound, we could then derive upper bounds on
$A(\bstar) - \bstar M (M-1)\gamma_S(\bstar)$.  However, in the limit of large
$M = \tau/\mylambda$, the density on $\Smat$ is effectively constrained to $(\bzero,
\btau)$ \cite{isit13, RoseMian16_1} which constrains $h(\Smat) \le M \log \tau$.  Then, since
$h(\Smat|\Tmat) = M (1-\log \mu)$ for exponential first-passage, \equat{Ibetamax1} produces
mutual information per token
\bee
\label{eq:IupperoverM}
\frac{I(\Svv;\Tmat)}{M}   \le   \log \tau - (1-\log \mu) 
  +   \log \left (1 +  {4 \myrho}  \right )
  -   \frac{\log M!}{M} \IEEEeqnarraynumspace
\eee
Application of Stirling's approximation for large $M$
\bee
\frac{\log M!}{M}
\approx
\log M - 1
\label{eq:stirling}
\eee
in combination with \equat{IupperoverM} produces our main theorem:
{\em \begin{theorem}{\bf An Upper Bound on the Asymptotic Capacity per Token, $C_q$:}
{\thmlabel{IupperoverM}}

For exponential passage with mean first-passage time $1/\mu$ and token emission intensity
$\mylambda$, an upper bound for the asymptotic capacity per token is given by
\bee
C_q
=
\max_{f_\Tmat(\cdot)}
\lim_{M \rightarrow \infty}
\frac{1}{M} I(\Svv;\Tmat)
\le
\log \left (\frac{1}{\myrho} + 4 \right )
\eee
where $\myrho = \frac{\mylambda}{\mymu}$.
\end{theorem}}
\begin{Proof}{\Thmref{IupperoverM}}
Substitution of \equat{stirling} and $\tau = M/\mylambda$ into \equat{IupperoverM} completes the proof.
\end{Proof}

\section{Discussion  \& Conclusion}
The timing channel \cite{bits-Qs, sundaresan1, sundaresan2, moewin16} is a building block upon which
the information theory of the identical molecule/token timing channel is built.  In this paper we
considered a version of the timing channel were a single emission is restricted to an interval
$[0,\tau]$ and we derived closed form expressions for the channel capacity under exponential
first-passage as well as the optimal input (emission) distribution.  We also established that unlike
for the mean-constrained channel, exponential first-passage is {\em not} the worst case corruption.

Building block though the single emission channel is, the identical molecule timing channel differs
from previous models because which emission corresponds to which arrival is ambiguous expressly
because travel time from sender to receiver is random and the molecules are identical.  This
ambiguity is captured by a quantity we define as the ``ordering entropy'' $H(\Omega|\Svv,\Tmat)$ and
understanding its properties is critical to understanding the capacity of not only the molecular
timing channel, but also channels where tokens/molecules may themselves carry information payloads
-- portions of messages to be strung together at the receiver \cite{RoseMian16_1}.

In the Part-I companion to this paper \cite{RoseMian16_1}, we carefully explored the information
theory formulation of the problem to establish that the usual information $I(\Smat;\Tmat)$ is indeed
the proper measure of information flow over this channel and its relationship to
$H(\Omega|\Svv,\Tmat)$.  In this paper, Part-II, we carefully explored the properties of
$H(\Omega|\Svv,\Tmat)$, showing how it can be calculated without deriving full order distributions
and deriving closed form expressions for cases where the emission times $\Tmat$ are i.i.d. random
variables.  We then derived closed form expressions for the special cases of the input distribution
being that which achieves capacity for the mean-constrained and the deadline-constrained timing
channel with exponential first-passage and the asymptotic behavior $\lim_{M \rightarrow \infty}
H(\Omega|\Svv,\Tmat)$.  Our understanding of $H(\Omega|\Svv,\Tmat)$ then allowed derivation of lower
bounds on timing channel capacity for exponential first passage (Part-I, Theorem 14) and here in
Part-II, an upper bound for the molecular timing channel capacity.

Although the machinery necessary to consider a mean-constrained version of the identical token
timing channel was derived, capacity results were not pursued owing to our inability to derive an
appropriate sequential channel use model with asymptotic independence.  However, if physically
parallel channels were used (so as to avoid corruption of one channel by arrivals from another), the
results of \cite{bits-Qs} combined with \Thmref{HOmaxmean} might be used to derive upper and lower
bounds analogous to those provided here and in Part-I \cite{RoseMian16_1}.  This might prove
interesting since the mean-constraint seems analytically simpler than the deadline constraint with
respect to both the single-token entropy and capacity as well as the ordering entropy.

\section*{Acknowledgments}
Profound thanks are owed to A. Eckford, N. Farsad, S. Verd\'u and V. Poor for useful discussions and
guidance.  We are also extremely grateful to the editorial staff and the raft of especially careful
and helpful anonymous reviewers. This work was supported in part by NSF Grant CDI-0835592.

\bibliographystyle{unsrt}
\bibliography{MERGE11,sm_collection,molecule-rose,biologicalcommunication,biologicalcommunicationCHRIS,mian2,NIS,RosePriorStandalone,Arash,citations,sm_collection}

\begin{thebibliography}{10}

\bibitem{RoseMian16_1}
C.~Rose and I.S. Mian.
\newblock {Inscribed Matter Communication: Part I}.
\newblock {\em {IEEE Transactions Molecular, Biological and Multiscale
  Communication}}, 2016.
\newblock (revised, Nov. 21) ArXiv: http://arxiv.org/abs/1606.05023.

\bibitem{bits-Qs}
V.~Anantharam and S.~Verdu.
\newblock {Bits Through Queues}.
\newblock {\em IEEE Transactions on Information Theory}, 42(1):4--18, January
  1996.

\bibitem{sundaresan1}
R.~Sundaresan and S.~Verd\'u.
\newblock {Robust Decoding for Timing Channels}.
\newblock {\em IEEE Transactions on Information Theory}, 46(2), 2000.

\bibitem{sundaresan2}
R.~Sundaresan and S.~Verd\'u.
\newblock {Capacity of Queues Via Point-Process Channels}.
\newblock {\em IEEE Transactions on Information Theory}, 52(6), 2006.

\bibitem{cover}
T.M. Cover and J.A. Thomas.
\newblock {\em Elements of Information Theory}.
\newblock Wiley-Interscience, 1991.

\bibitem{farsad_isit16}
N.~Farsad, Y.~Murin, A.~W. Eckford, and A.~Goldsmith.
\newblock {On the Capacity of Diffusion-Based Molecular Timing Channels}.
\newblock In {\em IEEE International Symposium on Information Theory 2016},
  pages 1023--1027, July 2016.

\bibitem{farsadIT16}
N.~Farsad, Y.~Murin, A.~W. Eckford, and A.~Goldsmith.
\newblock {Capacity Limits of Diffusion-Based Molecular Timing Channels}.
\newblock {\em {IEEE Transactions on Information Theory}}, 2016.
\newblock in preparation for submission.

\bibitem{isit11}
Y-L Tsai, C.~Rose, R.~Song, and I.S. Mian.
\newblock {An Additive Exponential Noise Channel with a Transmission Deadline}.
\newblock In {\em {IEEE International Symposium on Information Theory
  (ISIT'11)}}, pages 598--602, July 2011.

\bibitem{SonRos13}
C.~Rose, R.~Song, and I.S. Mian.
\newblock {Timing Channels with Multiple Identical Quanta}.
\newblock {\em {IEEE Transactions on Information Theory}}, 2013.
\newblock (in preparation, available: http://arxiv.org/abs/1208.1070).

\bibitem{Cover06}
T.~M. Cover and J.~A. Thomas.
\newblock Elements of information theory, 2006.

\bibitem{gallagerit}
R.G. Gallager.
\newblock {\em {Information Theory and Reliable Communication}}.
\newblock Wiley, 1968.

\bibitem{hild}
F.B. Hildebrand.
\newblock {\em Advanced Calculus for Applications}.
\newblock Prentice Hall, Englewood Cliffs, NJ, 1976.

\bibitem{isit13}
C.~Rose and I.S. Mian.
\newblock Signaling with identical tokens: Lower bounds with energy
  constraints.
\newblock In {\em IEEE International Symposium on Information Theory
  (ISIT'13)}, pages 1839--1843, July 2013.

\bibitem{moewin16}
G.~C. Ferrante, T.~Q.~S. Quek, and M.~Z. Win.
\newblock An achievable rate region for superposed timing channels.
\newblock In {\em 2016 IEEE International Symposium on Information Theory
  (ISIT)}, pages 365--369, July 2016.

\end{thebibliography}
\end{document}